\documentclass[12pt]{article}
\usepackage[a4paper, left=2cm,right=2cm]{geometry}
\usepackage{hyperref}
\usepackage{amsfonts} 
\usepackage{amsmath}
\usepackage{amssymb}
\usepackage{setspace}
\usepackage{color}

\usepackage[utf8,applemac]{inputenc}
\usepackage{tensor}
\usepackage{cite}
\usepackage{tikz}
\usepackage{graphicx,subfigure}
\usepackage[font=small, font+=it, format=hang, margin={1cm, 1cm}]{caption}
\graphicspath{{figure/}}

\usepackage{dcolumn}
\usepackage{bm}

\numberwithin{equation}{section}

\newcommand{\bea}{\begin{eqnarray}}
\newcommand{\eea}{\end{eqnarray}}
\newcommand{\be}{\begin{equation}}
\newcommand{\ee}{\end{equation}}
\def\nn{\nonumber}

\def\p{\partial}
\def\eps{\epsilon}

\newcommand{\cR}{\mathcal{R}}

\newcommand{\cM}{\mathcal{M}}
\newcommand{\cW}{\mathcal{W}}

\newcommand{\cE}{\mathcal{E}}

\renewcommand{\d}{\textrm{d}}

\begin{document}


\setcounter{tocdepth}{2}

\begin{titlepage}

\begin{flushright}\vspace{-3cm}
{\small
\today }\end{flushright}
\vspace{0.5cm}

\begin{center}

{{ \LARGE{\bf{Scalar Self-force for High Spin Black Holes\\ }}}}
\vspace{5mm}

\bigskip

\centerline{\large{\bf{Geoffrey Comp\`{e}re$^\dagger$\footnote{email: gcompere@ulb.ac.be}, Kwinten Fransen$^*$\footnote{email: kwinten.fransen@kuleuven.be}, Thomas Hertog$^*$\footnote{email:
thomas.hertog@kuleuven.be}}}}\vspace{2pt}
\centerline{\large{\bf{
and Yan Liu$^\dagger$\footnote{email: yliu6@ulb.ac.be}}}}

\vspace{2mm}
\normalsize
\bigskip\medskip
\textit{{}$^\dagger$ Centre for Gravitational Waves, Universit\'{e} Libre de Bruxelles, \\
 International Solvay Institutes, CP 231, B-1050 Brussels, Belgium}\\
\textit{{}$^*$ Centre for Gravitational Waves, Institute for Theoretical Physics, \\
KU Leuven, Celestijnenlaan 200D, 3001 Leuven, Belgium}

\vspace{15mm}

\begin{abstract}
\noindent
{
	We semianalytically investigate the scalar self-force experienced in the final stages of extreme mass ratio inspirals of nonspinning scalar particles into supermassive nearly extremal Kerr black holes. We exploit the near-horizon conformal symmetry to find the self-force for general corotating equatorial geodesics.  The angular component of the self-force is shown to be universal at leading order in the high spin limit. We verify that the energy and angular momentum losses of the scalar particle match with the asymptotic fluxes of scalar radiation. In particular, we relate the previously described  persistent oscillations in the asymptotic energy and angular momentum fluxes with the local self-force. Such oscillations arise from traveling waves that prevent the near-horizon and the asymptotic region to fully decouple in the extremal limit. Conformal invariance is therefore reduced to discrete scale invariance with associated logarithmic periodicity. 
}

\end{abstract}


\end{center}

\end{titlepage}

\tableofcontents

\section{Introduction}

Extreme mass ratio inspirals (EMRIs) of stellar mass black holes into supermassive black holes are one of the main target sources for the space-based gravitational wave detector LISA \cite{Audley:2017drz}. This is because they provide unique high-precision probes of strong gravity physics around black holes in the LISA frequency band thereby enabling new and powerful tests of general relativity and of our understanding of black holes. In addition the observation of EMRIs can provide information on the immediate environment of Milky Way like massive black holes \cite{AmaroSeoane:2007aw}. Although EMRI signals could potentially be in-band for a large number ($\sim 10^5$) of cycles, their signal to noise ratio (SNR) is expected to be low, implying that, save for lucky ``golden'' binaries, precise waveform models will be a prerequisite for both detection, parameter estimation and the extraction of interesting physics.

The large hierarchy of masses and associated inspiral timescales puts EMRIs beyond the reach of numerical relativity. On the other hand this makes these systems well-suited to a perturbative gravitational self-force approach, see \cite{Barack:2009ux,Poisson:2011nh,Barack:2018yvs} for modern reviews. Despite steady progress in the development and application of this self-force approach \cite{detweiler2003self, barack2007gravitational, barack2009gravitational, isoyama2014gravitational, merlin2016completion, fujita2017hamiltonian, barack2017time}, up to calculations of the first order gravitational self-force for generic bound geodesics in Kerr \cite{vandeMeent:2017bcc}, there are still several outstanding challenges. Notably, second order results are required to reach the desired accuracy for LISA and increasing the efficiency of computations will be indispensable to attain reasonable coverage of the large parameter space. Steps have been taken to address these issues \cite{pound2017nonlinear, van2018fast} but further significant efforts will be required to be able to fully exploit the science potential of EMRIs with LISA.
 
Progress in gravitational self-force calculations has typically been achieved by first starting with a scalar analogy of the full gravitational problem \cite{nasipak2019repeated} and with (quasi)circular orbits \cite{warburton2010self} around static central black holes \cite{diaz2004scalar}, and then progressively working to the actual gravitational case, to more general orbits and to rotating black holes. However, simplifications also occur for black holes rotating close to extremality. Indeed, the physics near the horizon of high-spin black holes is governed by an emergent $SL(2,\mathbb{R})$ conformal symmetry \cite{Bardeen:1999px}. Therefore, a high-spin approach potentially yields an alternative starting point to tackle the more general case, that is complementary to the usual static limit. In addition it will be a valuable approach to understand the high spin region in parameter space, which will be useful to confidently interpolate expected gravitational wave signals through the entire range of possible EMRIs. 
  
Moreover, extremal black holes have been paramount in advancing our theoretical understanding of black holes using holographic methods. While much progress has been done for supersymmetric black holes, extremal rotating black holes are more interesting with respect to astrophysics. The Thorne bound limits the angular momentum of black holes to be smaller than $99.8\%$ \cite{1974ApJ...191..507T}. However, there are known ways to exceed this bound, notably by the addition of magnetic fields \cite{2011A&A...532A..41S}. There is even observational evidence for highly spinning black holes \cite{Brenneman:2006hw, Brenneman:2013oba}, though the spin might not be high enough to describe the physics accurately with only the leading order in the high spin expansion. 

In this paper, we will further explore EMRIs with a central high spin black hole by making explicit and precise the suggestion of \cite{Hadar:2016vmk} to exploit the near-horizon conformal symmetry for self-force calculations at leading order in the high spin expansion. In principle, the self-force depends on the entire past of the trajectory which, due to the self-force, is non-geodesic. We shall however restrict ourselves to studying the self-force given a fixed geodesic trajectory. From this ``geodesic self-force'' for an entire collection of geodesics one could in principle make an osculating elements approximation to determine the inspiral. 

The paper is organized as follows. In Sec. \ref{sec:circular}, we first solve the problem of scalar self-force for circular orbits in the near-horizon geometry of near-extremal Kerr. In Sec. \ref{sec:SF}, we describe the procedure of exploiting the circular solution to find the self-force for more general equatorial orbits. We also provide an explicit example. We conclude in Sec. \ref{ccl}. In Appendix \ref{app:NHEK} we review basic aspects of the near-horizon extremal geometry and its relation to Kerr. We relegate the technical aspects of the mode-sum regularization method to Appendix \ref{app:reg}.

\section{Self-force for circular orbits}
\label{sec:circular}

\subsection{Scalar self-force}

We consider a scalar field $\Psi$ sourced by a scalar particle of charge density $\rho$ and charge $q$ on the worldline $\gamma$ determined by the trajectory $z^\mu(\tau)$ \footnote{Note that the mass $\mu$ of this scalar particle is not necessarily conserved but varies as 

\be
\frac{\d \mu}{\d \tau} = -\frac{\d z^{\alpha}}{\d \tau} F_{\alpha}.
\ee
This effect will become important when considering how the orbit evolves under the action of the self-force.},
\bea
\nabla_{\mu}\nabla^{\mu} \Psi = -4 \pi \rho,\qquad \rho = q \int_\gamma \frac{\delta^{(4)}(x^\mu - z^\mu(\tau))}{\sqrt{-g}} \d\tau .
\label{scalarwave}
\eea
The analogue of the Lorentz force on a particle of trajectory $z(\tau)$ and of scalar charge $q$ in an external complex Klein-Gordon field $\Psi$ is given by 
\bea
F_\mu(\tau) = \frac{q}{2} (\p_\mu \Psi)\vert_{z(\tau)} + c.c. \label{Lo}
\eea
The Lorentz force can be obtained from the stress-tensor as done in the Appendix A of \cite{Quinn:1999kj}. However, for a scalar field sourced on the worldline $z^\mu(\tau)$ itself, the expression \eqref{Lo} diverges and must be suitably regularized. The general result for the regular scalar self-force was first described in \cite{Quinn:2000wa}. According to the prescription of Detweiler and Whiting \cite{detweiler2003self}, one can separate the full retarded field associated to the worldline into a regular and a singular piece $\Psi = \Psi^R+\Psi^S$ in such a way that the force $F_{\mu}^R$ associated to $\Psi^R$ is completely regular and fully accounts for the self-force. A practical way to accomplish the subtraction of the singular component from the full force is by a mode-sum regularization \cite{Barack:1999wf}. In this method, one decomposes the fields in angular harmonics. Even if the full solution diverges, each individual mode is finite. The idea is therefore to simply subtract the divergent piece mode by mode. The divergent piece itself can be found by a local expansion around the source, which is known for certain types of motion in a Kerr background \cite{Barack:2002mh}. The purpose of this section is to derive a formula for the self-force on circular orbits in the near-horizon regions of near-extremal Kerr, which we refer to as NHEK or near-NHEK. For a review of the geometry of these regions, we refer the reader to \cite{Compere:2017hsi,Compere:2018aar} and to Appendix \ref{app:NHEK}. 

\subsection{NHEK circular orbits}

\subsubsection{Orbit}

The central black hole is assumed to be described by the Kerr geometry with near-extremality parameter $\lambda = \sqrt{1-J^2/M^4}$, with mass $M$ and angular momentum $J\equiv aM$.  In the terminology of \cite{Compere:2017hsi} and consistently with the notations of Appendix \ref{app:NHEK} for the Poincar\'e NHEK coordinates $(T,R,\theta, \Phi)$, the circular NHEK orbit ``Circular$_*$'' is defined as
\bea
R=R_0,\qquad \Phi = \tilde \Omega T,\qquad \tilde \Omega = -\frac{3}{4}R_0.
\eea
In relation to the full Kerr geometry, it depends only on $R_0 \lambda^{2/3}$, or $\hat x_0$ as defined by 
\bea
\hat x_0 \equiv  \frac{\hat r_0 - \hat r_+}{M} = R_0 \lambda^{2/3}\label{BLx}
\eea
where $\hat r_+$, $\hat r_0$ are respectively the Boyer-Lindquist radius of the horizon and the orbit,  and $R_0$ is the NHEK radius.

\subsubsection{Scalar waves}

Circular orbits emit scalar waves with a single frequency $\Omega = m \tilde \Omega$ for each azimuthal mode number $m$. The scalar solution of \eqref{scalarwave} in NHEK for a source on the circular orbit is given by
\bea
\Psi = \sum_{\hat l,m} R_{\hat lm\tilde \Omega}(R) S_{\hat l m}(\theta)e^{i m (\Phi - \tilde \Omega T)},
\eea
where $S_{\hat l m}$ are scalar spheroidal harmonics and $R_{\hat lm\tilde \Omega}(R)$ is, away from the source at $R=R_0$, a linear combination of the independent solutions to the homogeneous radial wave equation
\begin{eqnarray}
\cM^{\text{D}}_{\hat l m \Omega}(R) &=&  M_{im,h-1/2}(\frac{-2i \Omega}{R}), \quad \cM^{\text{D}}_{\hat l m 0}(R) = R^{-h}\\
\cW^{\text{in}}_{\hat l m \Omega}(R) &=& W_{im,h-1/2}(\frac{-2i \Omega}{R}), \quad \cW^{\text{in}}_{\hat l m 0}(R) = R^{h-1}
\end{eqnarray}
Here, the conformal weight $h$ will be given in terms of the spheroidal eigenvalues in \eqref{eqn:h}. The first set of modes obey Dirichlet boundary conditions at $R \rightarrow \infty$ which is the matching region with the asymptotically flat spacetime, $\cM^{\text{D}} \mapsto (-2 i \Omega)^h R^{-h}(1+O(R^{-1}))$ while the second set of modes obey ingoing boundary conditions at the horizon $R \rightarrow 0$, $\cW^{\text{in}} \mapsto (-2 i \Omega)^{im} R^{-i m}e^{i \Omega/R} $. We have included the degenerate $\Omega=0$ case to make the NHEK description self-sufficient but near-NHEK corrections are in principle necessary here to properly resolve what happens.  We use hatted labels for spheroidal modes and we save the unhatted labels for spherical modes, following the convention of \cite{warburton2011self}. The separation constants $\cE_{\hat lm}$ are determined by the angular eigenvalue problem whose solutions are the spheroidal harmonics (which are the spheroidal Legendre functions multiplied by $e^{im \Phi}$)
\be
\frac{1}{\sin{\theta}} \frac{\d}{\d \theta}(\sin{\theta} \frac{\d S_{\hat l m}}{\d \theta}) + [ \frac{m^2}{4} \cos^2{\theta}-\frac{m^2}{\sin^2{\theta}}+ \cE_{\hat l m} ] S_{\hat lm} = 0.  	
\label{eqn:extAngularTeukolsy}
\ee
We adopt the normalization 
\be
\int^{\pi}_0 |S_{\hat lm}(\theta)|^2 \sin{\theta} \d \theta = 1. 
\ee
For $m \neq 0$, the explicit solution reads as\footnote{The case $m=0$ can be solved separately. However, it is degenerate in the NHEK geometry since $\Omega=0$ indicates that $O(\lambda^{1/3})$ corrections are relevant, which are described by the near-NHEK geometry.}
\bea
R_{\hat lm\tilde \Omega}(R) &=& \frac{q S_{\hat l m}(\frac{\pi}{2})}{\sqrt{3}M} \frac{\Gamma(h-im)}{ i m  \Gamma(2h)}\Big( \mathcal M^{\text{D}}_0 \Theta(R_0-R)\mathcal W_{\hat l m \Omega}^{\text{in}}(R)  +  \mathcal W^{\text{in}}_0 (\Theta (R- R_0) \mathcal M_{\hat l m \Omega}^{\text{D}}(R) \nonumber \\
&& +Y(\lambda^{2/3} \Omega) \mathcal W_{\hat l m \Omega}^{\text{in}}(R))\Big)\label{solR}
\eea
where 
\bea
\mathcal M^{\text{D}}_0 & \equiv  & M^{\text{D}}_{\hat l m \Omega}(R_0) =  M_{im,h-\frac{1}{2}}(\frac{3}{2}im),\\
\mathcal W^{\text{in}}_0 & \equiv  &W^{\text{in}}_{\hat l m \Omega}(R_0) = W_{im,h-\frac{1}{2}}(\frac{3}{2}im), \\
Y(\lambda^{2/3} \Omega)&=& \frac{\Gamma(1-h-im)}{\Gamma(1-2h) (k^{-1}_2(-i \lambda^{2/3} \Omega)^{1-2h}-1)}, \label{eqn:Y}\\
k_2 &=& (-2 i m)^{2h-1} \frac{\Gamma(1-2h)^2}{\Gamma(2h-1)^2}\frac{\Gamma(h- i m)^2}{\Gamma(1-h-im)^2}.
\eea
The outgoing or ingoing character of a single mode is determined by the sign of $m$. We adopt the convention for the conformal weight $h \equiv h_{lm}$,
\bea
h &=& \left\{ \begin{array}{l} \frac{1}{2} +\frac{1}{2} (\sqrt{\eta_{lm}})^*\qquad \text{for}\qquad m \geq 0,\\
\frac{1}{2} +\frac{1}{2} \sqrt{\eta_{lm}}\qquad \text{for}\qquad m < 0 \end{array} \right. \label{eqn:h}
\eea
where $\eta_{lm} = 1-7 m^2 + 4 \mathcal E_{lm}$. This allows for a coherent boundary condition between positive and negative $m$ modes for the traveling waves, i.e. when $\eta_{lm}$ is negative and $h-1/2$ becomes imaginary. These modes occur when the ratio $\frac{m}{l+1/2}$ exceeds a critical ratio of approximately $0.74$ \cite{Yang:2012pj}. It also turns out to lead to $k_2$ exponentially suppressed for high $|m|$. The value of $k_2$ will play a critical role for the traveling wave modes since $\lambda^{2h-1}$ in \eqref{eqn:Y} is then no longer parametrically small but rather oscillatory. Nevertheless, $|k_2 \lambda^{2h-1}|$ will still be small in practice as for such modes $k_2 \lesssim 10^{-5}$. Our convention \eqref{eqn:h} differs from \cite{Yang:2012pj,Porfyriadis:2014fja,Compere:2017hsi}.

The scalar field with Dirichlet boundary conditions in NHEK is scale invariant and therefore does not depend upon $R_0$. This is why the first line of \eqref{solR} does not depend upon $R_0$. However, the boundary conditions that relate the asymptotic NHEK region to the asymptotically flat region break scale invariance. The homogeneous solution which has to be added to obey the asymptotically flat outgoing boundary conditions therefore depend upon the parameter $\hat x_0 = R_0 \lambda^{2/3}$ through the $Y$ coefficient in \eqref{solR}. 

We could also rewrite the solution in terms of the ingoing mode at the horizon $\cW^{\text{in}}_{\hat l m \Omega}(R)$ and the up mode which is outgoing at null infinity in the asymptotically flat region 
\bea
\cW^{\text{up}}_{\hat l m \Omega}(R ; \hat x_0)& \equiv & \cM^{\text{D}}_{\hat l m \Omega}(R) + Y(\lambda^{2/3} \Omega) \cW^{\text{in}}_{\hat l m \Omega}(R).
\eea
For $m \neq 0$, the explicit solution reads as
\be
R_{\hat lm\tilde \Omega}(R) = \frac{q S_{\hat l m}(\frac{\pi}{2})}{\sqrt{3}M} \frac{\Gamma(h-im)}{ i m  \Gamma(2h)}\Big( \mathcal W^{\text{up}}_0(\hat x_0) \Theta(R_0-R)\mathcal W_{\hat l m \Omega}^{\text{in}}(R)  +  \mathcal W^{\text{in}}_0 \Theta (R- R_0) \mathcal W_{\hat l m \Omega}^{\text{up}}(R ; \hat x_0) \Big), \label{eqn:solR}
\ee
where $\mathcal W_0^{\text{up}} (\hat x_0) = \mathcal M^{\text{D}}_0 + Y(\lambda^{2/3} \Omega) \mathcal W^{\text{in}}_0$. For our purposes, however, it will prove convenient to keep the formulation of \eqref{solR}, the reason being that it separates the problem of boundary conditions breaking the symmetry of the background from the regularization of the local self-force divergence.  For axisymmetric modes $m=0$\footnote{This solution can be formally obtained from the $m \rightarrow 0$ limit of the $m \neq 0$ solution using $h > 0$.}
\bea
R_{\hat l 0 0}(R) &=& -\frac{\sqrt{3}q S_{\hat l0}(\pi/2)}{2 M}  \frac{R_0}{1-2h}
(R_0^{-h} \Theta(R_0-R) R^{-1+h} + R_0^{-1+h} \Theta(R-R_0) R^{-h}) 
\label{eqn:degscalarNHEKsolution}
\eea
where $h_{l0} = 1+l$. In this case, the behavior $R^{-h}$ extends to the asymptotically flat region, which simply corresponds to a multipole deformation.

\subsubsection{Self-force}

\begin{figure}[!hbt]
	\centering
	\subfigure{
		\includegraphics[width=.45\textwidth]{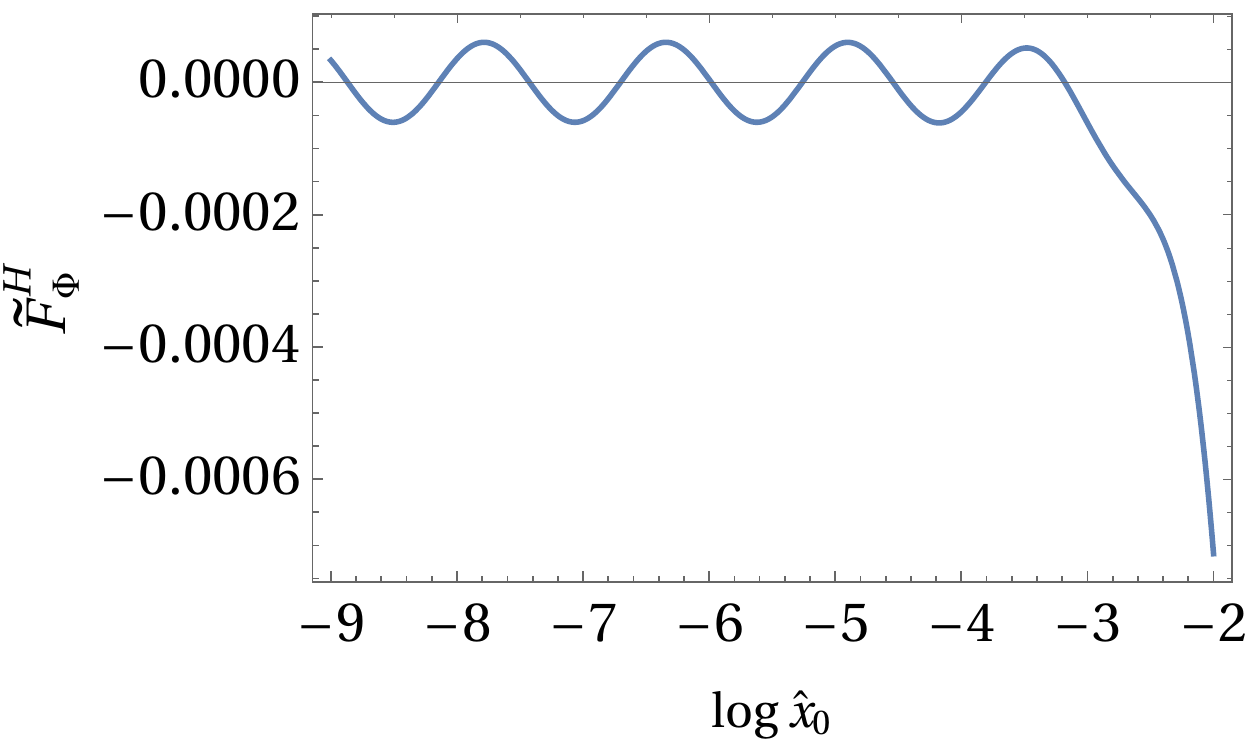}}
	\qquad
	\subfigure{
		\includegraphics[width=.45\textwidth]{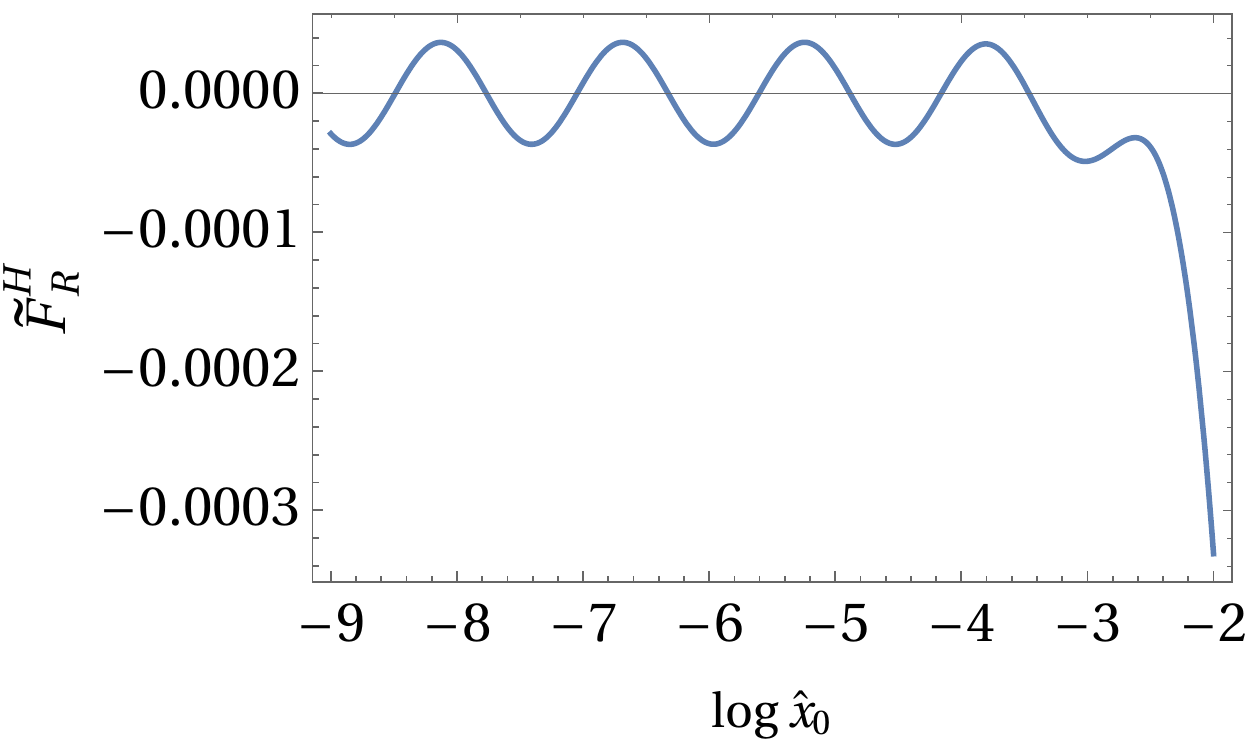}}
	\qquad
	\caption{$\tilde F_{\Phi}^H( \hat x_0)$ (left) and $\tilde F_R^H( \hat x_0)$ (right) both computed summing up to $\hat l = 30$.}
	\label{fig:FhomNHEK}
\end{figure}

Let us now turn our attention to the self-force. It is clear that, in the equatorial plane, $F_\theta = 0$ by symmetry. Also, $\mathcal L_\xi \Psi = 0$ for $\xi=\p_{\hat t}$, which translates into $F_T = -\tilde \Omega F_\Phi$. We shall therefore only compute $F_R$, the conservative part of the self-force, and $F_\Phi$, the dissipative part of the self-force. We define the adimensional quantities $\tilde F_R$, $\tilde F_{\Phi}$ as 
\bea
F_R = \frac{q^2}{M R_0} \tilde F_R( \hat x_0), \qquad F_\Phi = \frac{q^2}{M} \tilde F_\Phi( \hat x_0).
\eea
The computation  can be separated into two parts. First, we will obtain the self-force for the inhomogeneous problem with Dirichlet boundary conditions in NHEK, which requires a regularization. The answer for both  $\tilde F^I_R$ and $\tilde F^I_T$ will be a pure number which we determine numerically. We will then derive the self-force for an additional homogeneous perturbation, which does not require any regularization. Moreover, one mode turns out to dominate as $\lambda \to 0$ and therefore it can be understood easily in its analytic form. That part depends upon $ \hat x_0 = R_0 \lambda^{2/3}$. The final answer is 
\bea
\tilde F_R = \tilde F_R^{I} + \tilde F_R^H( \hat x_0), \label{eqn:NHEKHIsplitR} \\
\tilde F_\Phi = \tilde F_\Phi^{I} + \tilde F_\Phi^H( \hat x_0). \label{eqn:NHEKHIsplitPhi}
\eea
For the homogeneous part, we find
\bea
\tilde F_\Phi^H( \hat x_0) &=& \frac{1}{\sqrt{3}} \sum_{\hat l,m \neq 0} \Big( \frac{[S_{\hat lm}(\frac{\pi}{2}) \mathcal W^{\text{in}}_0]^2 }{(k_2)^{-1} (\frac{3im}{4}\hat x_0)^{1-2h}-1} \frac{\Gamma(h- im)\Gamma(1-h-im)}{\Gamma(2h)\Gamma(1-2h)} \Big),\label{N1}\\
\tilde F_R^H( \hat x_0) &=& -\frac{\sqrt{3}}{2} \sum_{\hat l,m\neq0 } \Big( \frac{[S_{\hat lm}(\frac{\pi}{2})]^2 \mathcal W^{\text{in}}_0\mathcal W^{\text{in}\prime}_0}{(k_2)^{-1} (\frac{3im}{4}\hat x_0)^{1-2h}-1} \frac{\Gamma(h- im)\Gamma(1-h-im)}{\Gamma(2h)\Gamma(1-2h)} \Big),\label{N2}
\eea
where $\mathcal W^{\text{in}\prime}_0 \equiv W^\prime_{im,h-\frac{1}{2}}(\frac{3}{2}im)$ and all other quantities were defined previously. We plot $\tilde{F}_R^H( \hat x_0)$ and $\tilde F_\Phi^H( \hat x_0)$ in Fig. \ref{fig:FhomNHEK}. Remark that, since $k_2 \ll 1$ for most modes, each term in the sum goes like $e^{(2h-1) \log \hat x_0}$. From Fig. \ref{fig:FhomNHEK}, it can be seen that in fact only, the $l=2$, $m=\pm2$ modes, dominate the low $\lambda$ behavior. This is the reason for the clean logarithmic oscillatory behavior of the self-force $F \propto \cos{(2\delta_{22} \log \hat x_0)}$ with $i\delta_{22}  = h_{22}-1/2$ in the low $\lambda$ limit as observed for the flux in \cite{Gralla:2015rpa} and, for the ISCO shift in \cite{vandeMeent:2016hel}. To compare with the latter, one can use the exact same functional form but simply using the appropriate $s=-2$ spin-weighted spheroidal harmonic eigenvalue in the definition of $h$ \eqref{eqn:h}. Although the existence of these low ``temperature'' oscillations have been seen previously in several observables, the question has been raised as to clarify their geometric origin \cite{vandeMeent:2016hel}, which to our best knowledge has not yet received a satisfying answer. Mathematically, these oscillations clearly arise from the existence of traveling waves that continue to connect the limit geometries as $\lambda \to 0$, i.e. the near-horizon decoupling is perturbatively unstable. This in turn breaks the scale invariance of the near-horizon in this limit to a discrete scale invariance characteristically associated to logarithmic periodicity \cite{hartnoll2016thermal}. What is particularly intriguing from the point of view of black hole physics is that this behavior is typical for systems with quenched disorder \cite{de2001exotic, weinrib1983critical}, an observation that seems to resonate with SYK-type approaches to black holes \cite{kitaev2015simple,sachdev1993gapless}. We will not speculate here but simply remark that it would be interesting to explore if such connections could have bearing on gravitational wave observables.    \\

For the inhomogeneous part, the application of \eqref{Lo} yields

\bea
\tilde F_\Phi^I &=&  \frac{1}{\sqrt{3}} \sum_{\hat l,m \neq 0} \Big( [S_{\hat lm}(\frac{\pi}{2})]^2 \frac{\Gamma(h-im)}{\Gamma(2h)} \mathcal M^{\text{D}}_0 \mathcal W^{\text{in}}_0 \Big),\nn \\
\tilde F_R^{I, +} &=& -\frac{\sqrt{3}}{2} \sum_{\hat l} \Big(\frac{[S_{\hat lm}(\frac{\pi}{2})]^2 h}{2h-1} + \sum_{m \neq 0} [S_{\hat lm}(\frac{\pi}{2})]^2 \frac{\Gamma(h-im)}{\Gamma(2h)} \mathcal M^{\text{D}}{}'_0 \mathcal W^{\text{in}}_0   \Big) ,\label{NHEKFI} \\
\tilde F_R^{I, -} &=& -\frac{\sqrt{3}}{2} \sum_{\hat l} \Big(\frac{[S_{\hat lm}(\frac{\pi}{2})]^2 (1-h)}{2h-1} + \sum_{m \neq 0} \frac{\Gamma(h-im)}{\Gamma(2h)} \mathcal M^{\text{D}}{}_0 \mathcal W^{\text{in}}{}'_0   \Big) ,\nn
\eea
where $\mathcal M^{\text{D}}{}'_0 = M^\prime_{im,h-\frac{1}{2}}(\frac{3}{2}im)$.

\begin{figure}[!hbt]
	\centering
	\includegraphics[width=.45\textwidth]{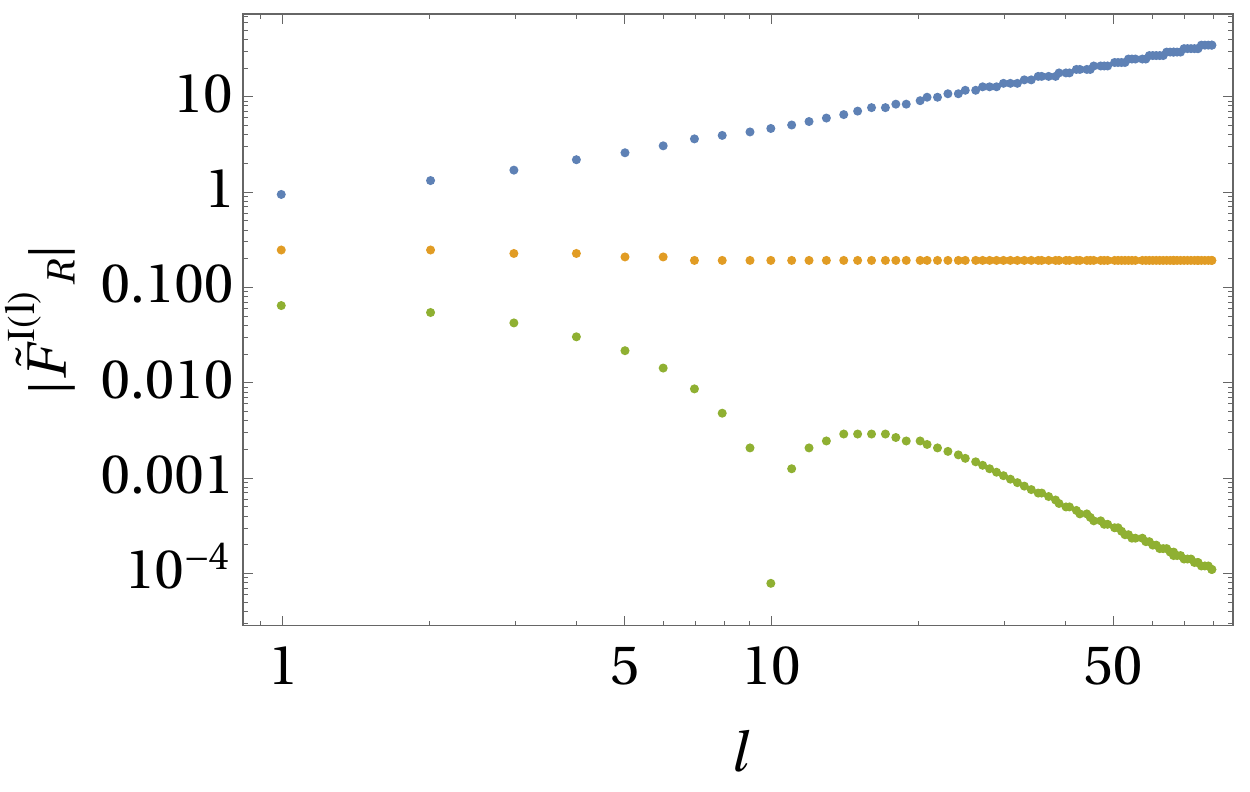}
	\caption{Different steps in the mode-sum regularization of $\tilde{F}_R$, starting with the unregularized l-modes (upper blue) and subtracting subsequently the analytically determined singular part linear in $l$ (middle orange) and both the linear and constant in $l$ singular contributions (lower green). }
	\label{fig:reg}
\end{figure}

It turns out that no regularization is needed for $\tilde F_\Phi^I$, such that this expression is fine as given, while regularization is required for $\tilde F_R^I$, as is true more generally for Kerr circular orbits \cite{warburton2010self}. Indeed, we note that the solution \eqref{solR} is continuous at $R =R_0$ while its radial derivative is not such that we needed to distinguish $F_R^{I, \pm}$ depending on how the limit $R \to R_0$ is taken. This distinction will disappear after regularization. We employ a mode sum regularization method \cite{Barack:1999wf} to regulate the radial part of the self-force. The details are relegated to Appendix \ref{app:reg} but the successive steps of substracting the dominant in $l$ singular pieces is shown in Fig. \ref{fig:reg}. Perfect cancellation is found between the retarded modes and the regularization, which are computed by a completely independent method.

\begin{figure}[!hbt]
	\centering
	\subfigure{
		\includegraphics[width=.44\textwidth]{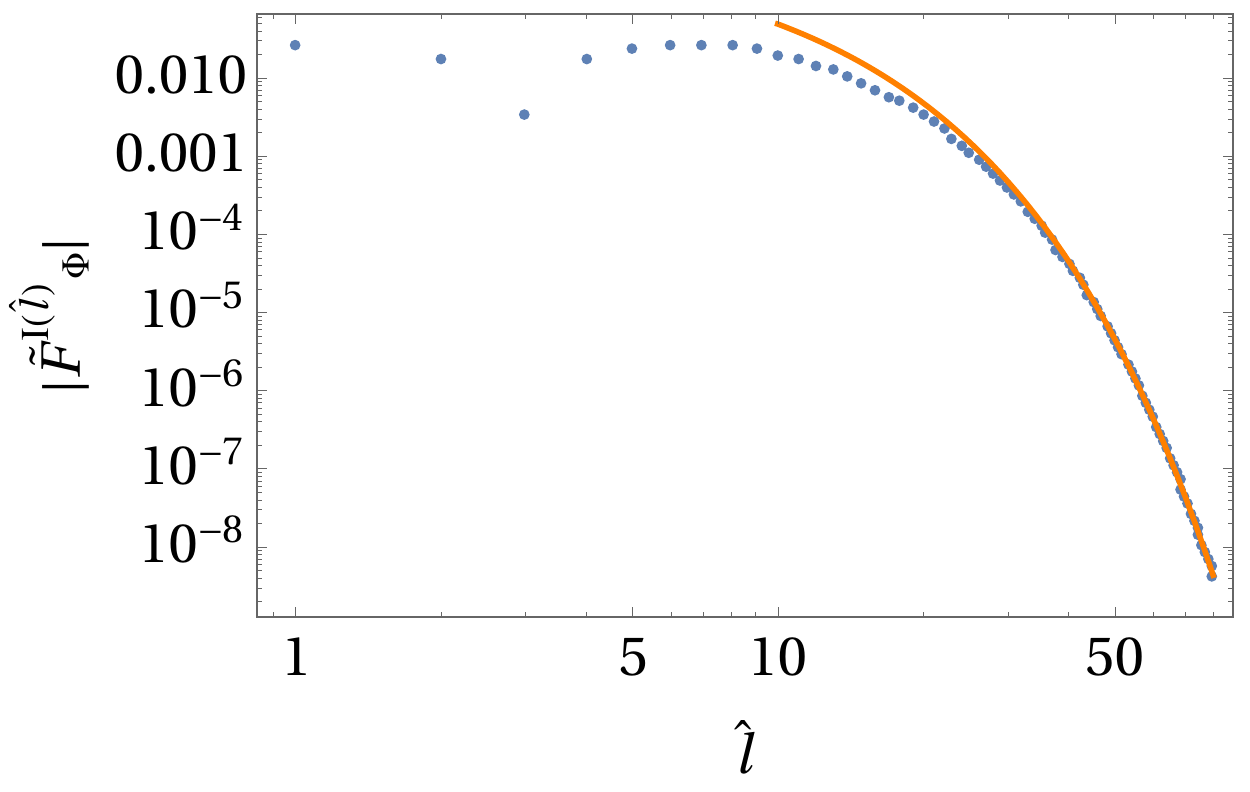}}
	\qquad
	\subfigure{
		\includegraphics[width=.44\textwidth]{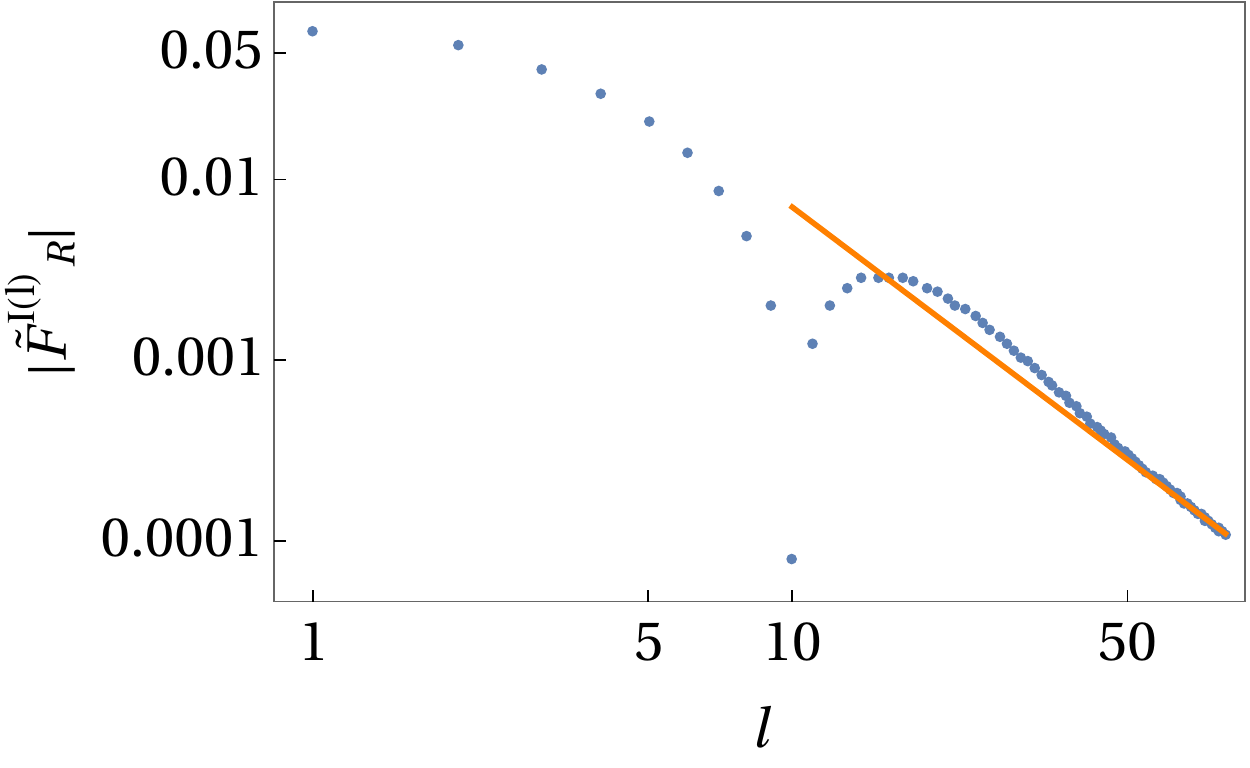}}
	\qquad
	\caption{$\tilde F_{\Phi}^I{}^{(\hat l)}$ (blue) with the best fit decaying exponential to the large $\hat l$ tail (orange) (left) and $\tilde F_R^I{}^{(l)}$ (blue) with the $l^{-2}$ tail behavior (orange) (right).}
	\label{fig:FinhomNHEK}
\end{figure}

The sums in \eqref{NHEKFI}, including the regularization we have just described, are performed numerically and therefore need to be cut off at finite $\hat{l}$. As is shown in Fig. \ref{fig:FinhomNHEK}, however, the large $\hat l$ contributions to $\tilde{F}^I_{\Phi}$ behave exponentially, and similarly the large $l$ (regularized) contributions to $\tilde{F}^I_{R}$ have an inverse quadratic behavior. This allows us to at least approximate these contributions beyond the numerical cutoff. Schematically, write $\tilde{F}^I_X = \sum^{l_{\text{cutoff}}}_{l=0} \tilde{F}^{(l)}_X + \tilde{F}^{(\text{tail})}_X$ with $X = \Phi, R$ (appropriately using $\hat l$ or $l$) and $\tilde{F}^{(l)}_X$ is determined from \eqref{NHEKFI} (including regularization), already summing over $m$. We approximate $\tilde{F}^{(\text{tail})}_{\Phi} \approx \sum^{\infty}_{\hat l= \hat l_{\text{cutoff}}+1} a_0 e^{-a_1 \hat l}$, $\tilde{F}^{(\text{tail})}_{R} \approx \sum^{\infty}_{ l=l_{\text{cutoff}}+1} \frac{a_2}{l^2}$ with $a_i$ coefficients which we determine by a numerical fit, as shown in Fig. \ref{fig:FinhomNHEK}. Performing this procedure for $\hat l_{\text{cutoff}}=80$ (respectively $l_{\text{cutoff}}=80$) and accounting for the estimation of the tail one finds 
\bea
\tilde F_{\Phi}^{I} = -0.22501771,\label{FPhiI} \qquad \tilde F_{R}^{I} = -0.204.
\eea
The error estimates, i.e. the tail components in these calculations are given respectively by $\tilde{F}^{(\text{tail})}_{\Phi} \approx 1.7 \times 10^{-8}$ (with $\hat l_{\text{cutoff}} = 80$) and $\tilde{F}^{(\text{tail})}_{R} \approx 0.009$  (with $l_{\text{cutoff}} = 80$). 

Note that $|\tilde F_{\Phi}^{I}| \gg |\tilde F_{\Phi}^{H}|$ and $|\tilde F_{R}^{I}| \gg |\tilde F_{R}^{H}|$ which can be interpreted to mean that the Dirichlet boundary conditions capture well the interaction with the asymptotic flat space, even if the traveling modes do not entirely decouple, i.e. this hierarchy is not parametric in $\lambda \to 0$.

\subsubsection{Local force vs fluxes}

As a check on the dissipative part of the self-force, we compare the energy and angular momentum fluxes to the local work done by the self-force.  The energy and angular momentum fluxes are not independent for circular orbits, $ \hat \Omega \frac{\d \ell}{\d \hat t} =  \frac{\d \hat E}{\d \hat t}$, with here $\hat \Omega \to \Omega_{\text{ext}} = \frac{1}{2 M}$ since we only look at the leading high spin limit. Hence, we can restrict to comparing the energy flux. Now it follows from
\be
\frac{\d \ell}{\d \hat t} = \frac{\sqrt{3} R_0 \lambda^{2/3}}{4} F_{\Phi}
\ee
that
\be
\frac{\d \hat E}{\d \hat t} = \frac{\sqrt{3} R_0 \lambda^{2/3}}{8 M} F_{\Phi} \approx - (R_0 \lambda^{2/3})\frac{q^2}{M^2} 0.0487.\label{res1}
\ee

To compare to a independent flux calculation we can use the results of \cite{Gralla:2015rpa} to get to leading order
\be
-\frac{\d \hat E}{\d \hat t} = (C_{\infty} + C_{H})(\lambda^{2/3}R_0) \approx 0.0487(\lambda^{2/3}R_0)\frac{q^2}{M^2}\label{res2}
\ee
using
\bea
C_{\infty} &\approx& 0.0745 \frac{q^2}{M^2}, \\
C_{H} &\approx& -0.0258 \frac{q^2}{M^2}.
\eea
We find perfect agreement between \eqref{res1} and \eqref{res2}.

\subsection{Near-NHEK circular orbits}

There are two conformal classes of orbits in NHEK \cite{Compere:2017hsi}. The first has the NHEK circular orbit as representative while the second has the near-NHEK circular orbit as representative. It is therefore crucial to study the circular near-NHEK orbit in order to describe the self-force on equatorial geodesics in NHEK. 

\subsubsection{Orbit}

The near-NHEK coordinates $(t,r,\theta,\phi)$ are defined in Appendix \ref{app:NHEK}.  In the terminology of \cite{Compere:2017hsi}, the circular near-NHEK orbit Circular$(\ell)$ is  defined as
\bea
r=r_0,\qquad \phi = \tilde \omega t,\qquad \tilde \omega = -\frac{3}{4}(r_0+\kappa).\label{nearNHEKcirc}
\eea
Here $\ell = J/\mu$ is the specific angular momentum of the orbit in near-NHEK that is associated to the Killing vector $\partial_{\phi}$. It relates to $\kappa_0 = \frac{\kappa}{r_0}$ as
\be
\kappa_0 = (\frac{2 \ell}{\sqrt{3(\ell^2-\ell^2_*)}}-1)^{-1}.
\label{def:kappa0}
\ee
In terms of the Boyer-Lindquist radius \eqref{BLx}, we have
\bea
\kappa_0 = \frac{\lambda}{\hat x_0} = \frac{\lambda^{1/3}}{R_0}.
\eea
The circular orbit is defined for a range of radii between the ISCO at $\ell=\ell_*=\frac{2}{\sqrt{3}}M$ and the light-ring at $\ell \to \infty$: $\kappa_0 $ lies in the range $0<\kappa_0 < \kappa_{l.r.}$  where $\kappa_{l.r.}  \equiv (\frac{2}{\sqrt{3}}-1)^{-1} \approx 6.46$ is the positive root of $3+6 \kappa_0-\kappa_0^2=0$.  Naturally, being beyond the innermost stable circular orbit, these trajectories are unstable. Although they can perhaps capture some effects of a zoom in certain zoom-whirl orbits \cite{healy2009zoom}, or locally approximate ``slow'' plunges, our main interest will be in using them to construct the self-force for equatorial orbits from conformal transformations.

\subsubsection{Scalar wave}

Circular orbits in near-NHEK emit scalar waves with frequency $\omega = m \tilde \omega$. The scalar solution to the wave equation on \eqref{scalarwave} on near-NHEK for the circular orbit is given by 
\bea
\Psi = \sum_{\hat l,m} R_{\hat lm\tilde \omega}(r) S_{\hat l m}(\theta)e^{i m (\phi - \tilde \omega T)}
\eea
where $S_{\hat l m}$ are again scalar spheroidal harmonics and, away from the source, $R_{\hat lm\tilde \omega}(r)$ is a linear combination of the independent solutions\footnote{Note that our normalization differs from \cite{Compere:2017hsi}.}
\begin{eqnarray}
\cR_{\hat lm\tilde \omega}^{\text{D}}(r) = (\frac{r}{2 \kappa})^{-h}(\frac{2\kappa}{r}+1)^{i(\frac{n}{2}-m)} {}_2F_1(h-im,h-im+in,2h,-\frac{2 \kappa}{r}),\\
\cR_{\hat lm\tilde \omega}^{\text{in}}(r) = (\frac{r}{2 \kappa})^{-in/2}(\frac{r}{2\kappa}+1)^{i(\frac{n}{2}-m)} {}_2F_1(h-im,1-h-im,1-in,-\frac{r}{2 \kappa}),
\end{eqnarray}
with $n \equiv m+\frac{{\omega}}{\kappa}$. The explicit solution for the scalar charge on a circular orbit is given by
\bea
R_{lm \tilde{\omega}}(r) &=& -\frac{q \sqrt{3+6\kappa_0-\kappa^2_0}}{4 M \kappa_0} S_{\hat l m}(\frac{\pi}{2}) \frac{\Gamma(h-i m) \Gamma(h-i (n-m))}{(1-2h)\Gamma(2h-1) \Gamma(1-i n)} \nn \\ &\times& \Big( \cR^{\text{D}}_0 \Theta(r_0-r) \cR^{\text{in}}(r) + \cR^{\text{in}}_0 (\Theta(r-r_0) \cR^{\text{D}}(r)+   Y (m \tilde{\omega},\lambda) \cR^{\text{in}}(r)) \Big)
\label{eqn:nearRadialSolution}
\eea
where 
\bea
\cR^{\text{D}}_0(\kappa_0) &\equiv& \cR^{\text{D}}(r_0)    \\
\cR^{\text{in}}_0(\kappa_0) &\equiv& \cR^{\text{in}}(r_0)    \\
Y(\omega,\lambda)&=&   \frac{\Gamma(h-i(n-m)) \Gamma(1-h-im)}{ \Gamma(1-2h) \Gamma(1-in) \left(k_2^{-1} \lambda^{1-2h} - \frac{ \Gamma(h-in+im)}{ \Gamma(1-h-in+im)} \right)}\label{Ynear}
\eea
The Teukolsky perturbation for Dirichlet boundary conditions only depends upon $\kappa_0$, not on $r_0$ or $\kappa$ independently.  Note that $Y=0$ for $m=0$ although the expression \eqref{Ynear} is ambiguous in that case. For the circular orbit this corresponds also to $\omega = m \tilde{\omega} = 0$ and, in fact, in this case, the near-NHEK solutions become exact solutions to the full Kerr scalar wave equation. The boundary behavior $\sim r^{-h} = r^{-l-1}$ is then the choice for a multipole deformation in asymptotically flat space as opposed to the tidal driving represented by $\sim r^{l}$ and is in particular immediately the right one at asymptotically flat space. From the point of view of the ordinary differential equation governing the radial direction, the loss of the radiation zone in this limit turns the irregular singular point at infinity for the full Kerr wave equation into a regular singular point. This can be fully captured by the near-zone, thus effecting this simplification.

\subsubsection{Self-force}
\label{sec:nearNHEKSF}

As in the case of NHEK, define
\bea
F_r &=& \frac{q^2}{Mr_0} \tilde F_r( \kappa_0 ; \lambda), \\
F_\phi &=& \frac{q^2}{M} \tilde F_\phi(\kappa_0 ; \lambda)
\eea

We again split the contribution from Dirichlet boundary conditions from the contribution from the homogeneous solution that restores the correct asymptotically flat boundary condition
\bea
\tilde F_r = \tilde F_r^{I}( \kappa_0) + \tilde F_r^H( \kappa_0 ; \lambda), \label{eqn:nearNHEKHIsplitr} \\
\tilde F_\phi = \tilde F_\phi^{I}( \kappa_0 ) + \tilde F_\phi^H( \kappa_0 ; \lambda). \label{eqn:nearNHEKHIsplitphi}
\eea

\begin{figure}[!hbt]
	\centering
	\subfigure{
		\includegraphics[width=.45\textwidth]{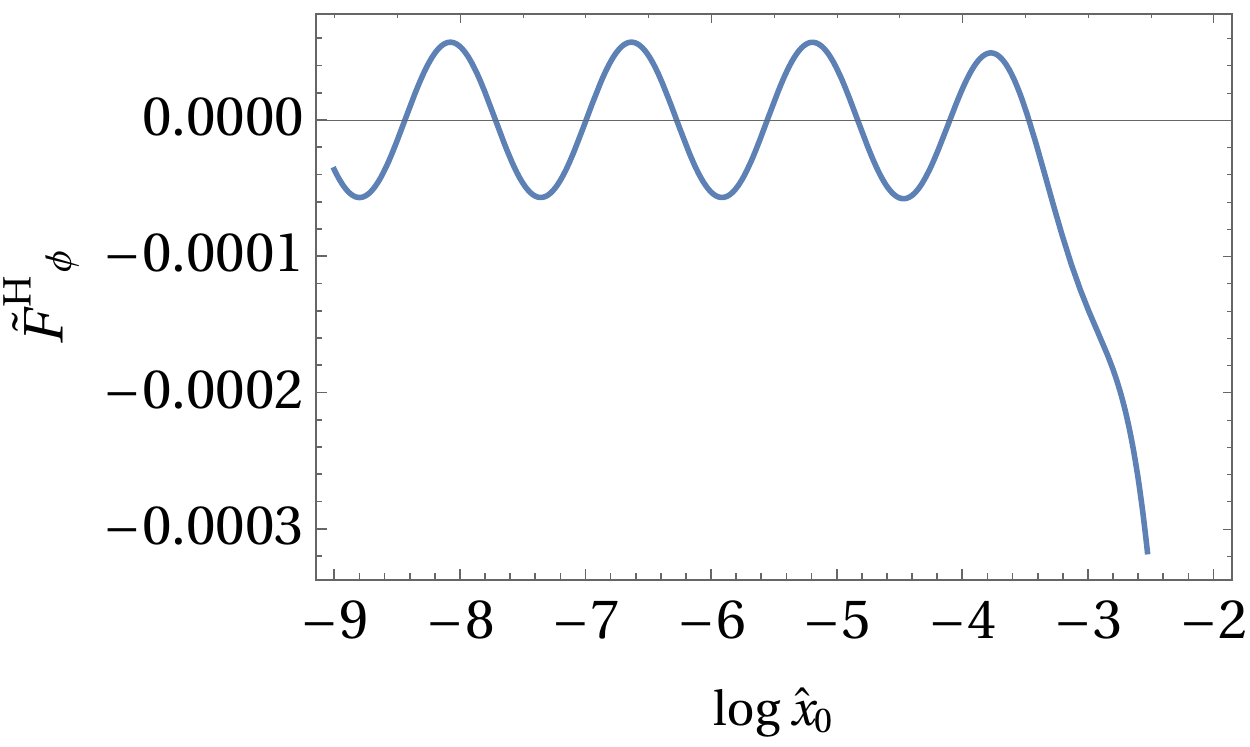}}
	\qquad
	\subfigure{
		\includegraphics[width=.45\textwidth]{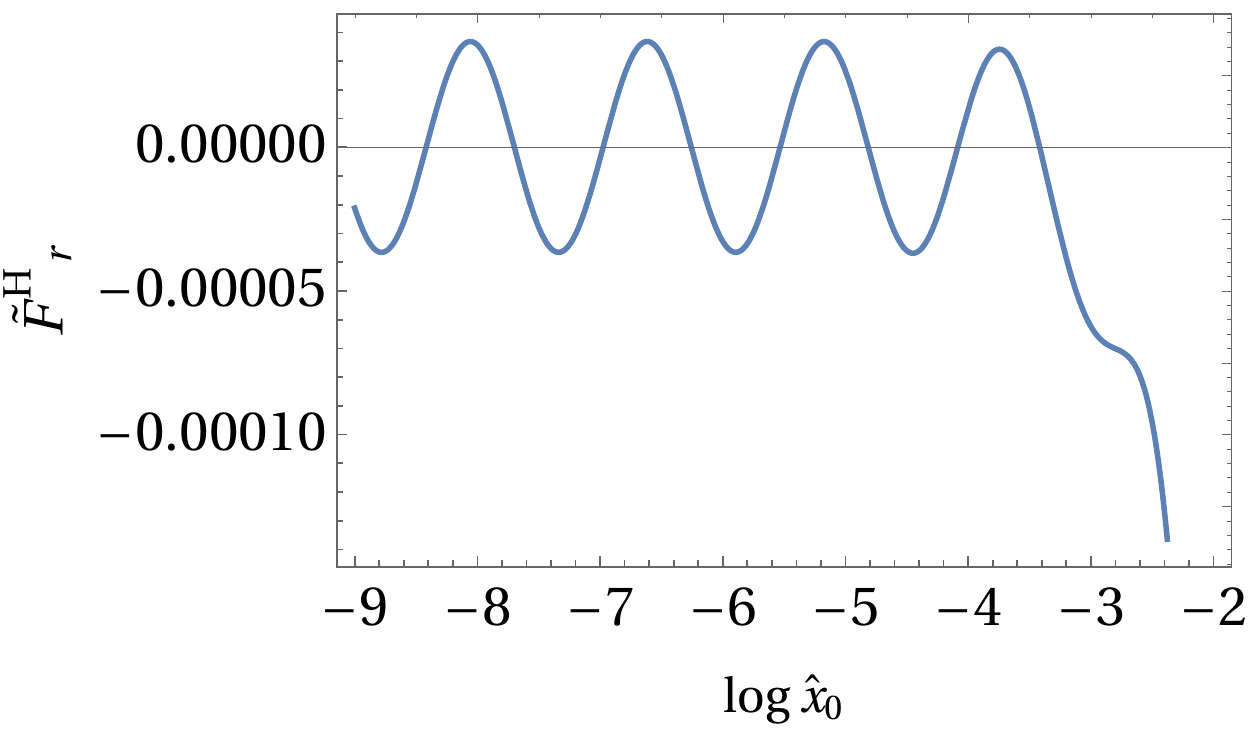}}
	\qquad
	\caption{$\tilde F_{\phi}^H(\hat x_0)$ (left) and $\tilde F_r^H( \hat x_0)$ (right) for $\kappa_0 = 1$ both computed summing up to $\hat l = 30$.}
	\label{fig:FhomnearNHEK}
\end{figure}

For the homogeneous part we find 
\be
\tilde F_{\phi}^H( \hat x_0) =   \sum_{\hat l, m \neq 0} \Big ( \frac{im \sqrt{3+6\kappa_0-\kappa^2_0} [S_{\hat lm}(\frac{\pi}{2}) \cR^{\text{in}}_0]^2}{k_2^{-1 }\lambda^{1-2h} - \frac{ \Gamma(h-in+im)}{ \Gamma(1-h-in+im)}}  \frac{\Gamma(1-h-im) \Gamma(h-i m) \Gamma(h-i (n-m))^2}{4 \kappa_0 \Gamma(2h)\Gamma(1-2h) \Gamma(1-i n)^2} \Big),
\label{eqn:FphiH}
\ee
\be
\tilde F_r^H( \hat x_0) =   \sum_{\hat l, m} \Big ( \frac{\sqrt{3+6\kappa_0-\kappa^2_0} [S_{\hat lm}(\frac{\pi}{2})]^2 \cR'^{\text{in}}_0 \cR^{\text{in}}_0 }{k_2^{-1 }\lambda^{1-2h} - \frac{ \Gamma(h-in+im)}{ \Gamma(1-h-in+im)}}  \frac{\Gamma(1-h-im) \Gamma(h-i m) \Gamma(h-i (n-m))^2}{4 \kappa^2_0 \Gamma(2h)\Gamma(1-2h) \Gamma(1-i n)^2} \Big),
\label{eqn:FrH}
\ee
where $\cR'^{\text{in}}{}_0(\kappa_0) \equiv \frac{\d \cR^{\text{in}}}{\d (r/\kappa)}(r_0)$. This is illustrated in Fig. \ref{fig:FhomnearNHEK} for $\kappa_0=1$ and can be seen to be qualitatively similar to the NHEK case. 

\begin{figure}[!hbt]
	\centering
	\subfigure{
		\includegraphics[width=.45\textwidth]{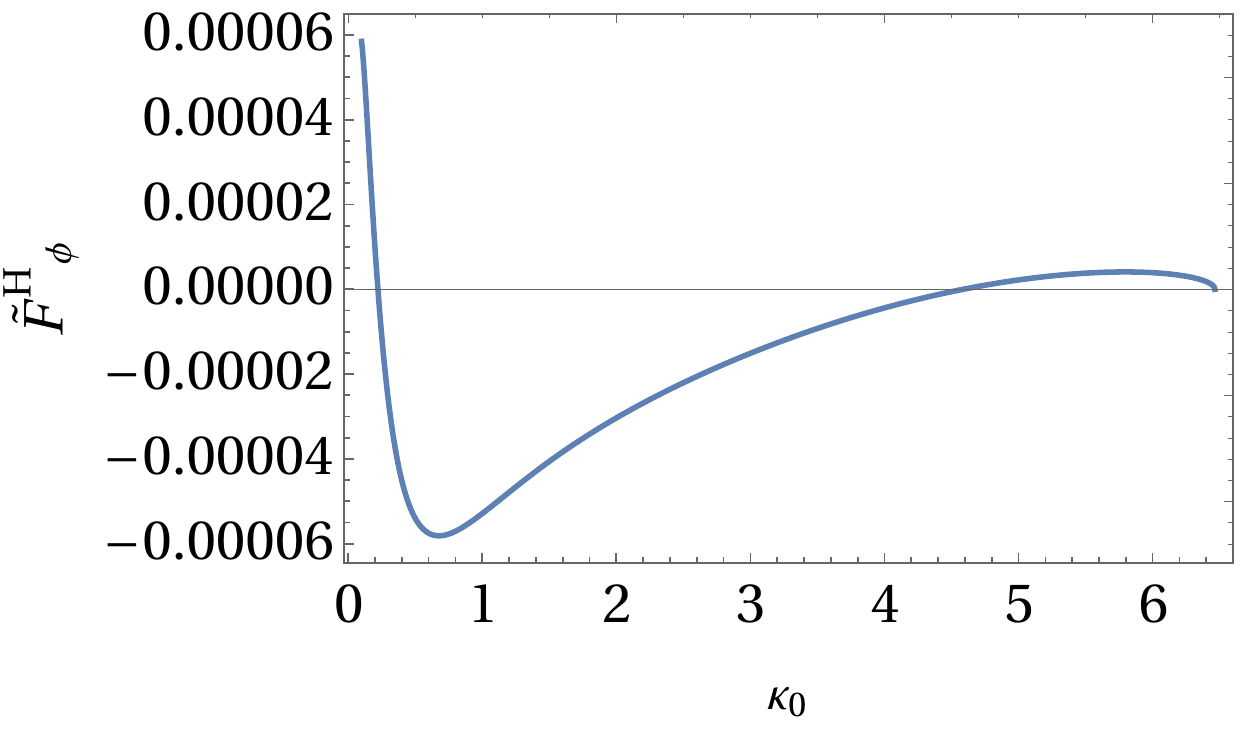}}
	\qquad
	\subfigure{
		\includegraphics[width=.45\textwidth]{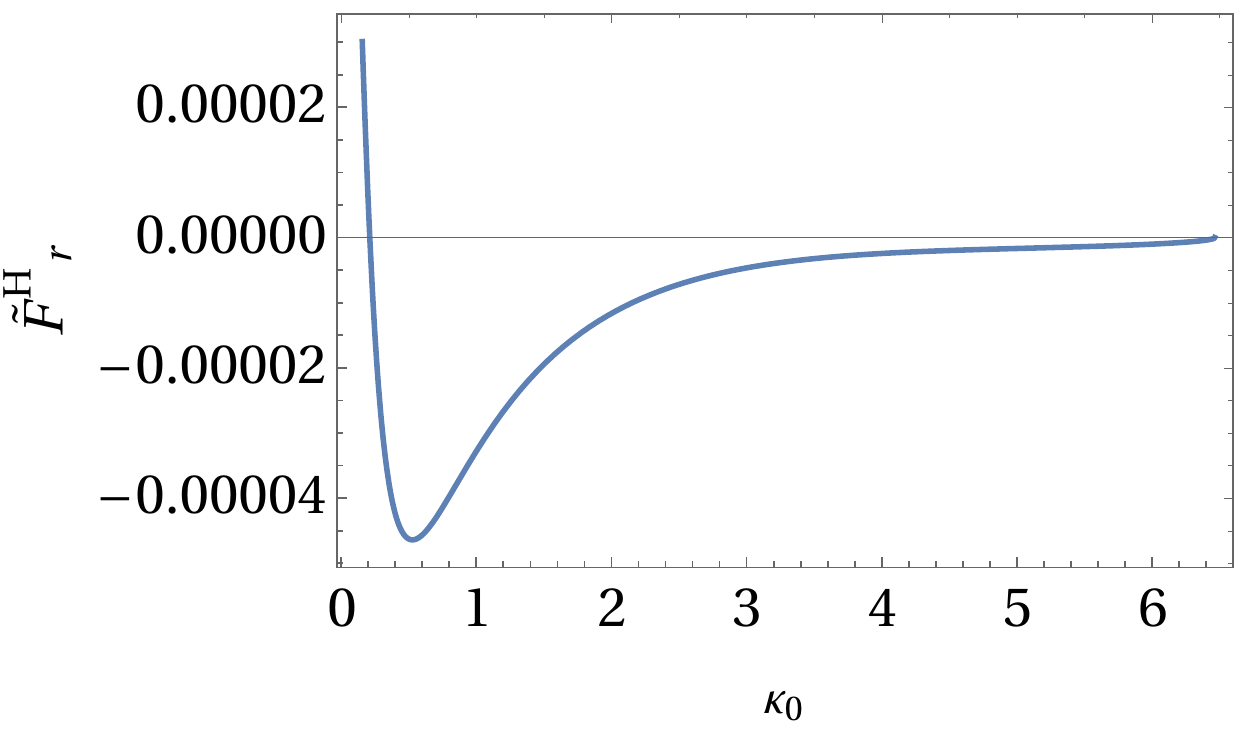}}
	\qquad
	\caption{$\tilde F_{\phi}^H(\kappa_0;10^{-6})$ (left) and $\tilde F_r^H(\kappa_0;10^{-6})$ (right) both computed summing up to $\hat l = 30$.}
	\label{fig:FhomnearNHEKkappa}
\end{figure}

Conversely, fixing $\lambda = 10^{-6}$, and varying $\kappa_0$ gives the image illustrated in Fig. \ref{fig:FhomnearNHEKkappa}, although this picture is less representative of the parameter space as it can vary more strongly with the chosen value of $\lambda$. 

A generic feature is that both self-force asymptote to zero as one approaches the light-ring, $\kappa_0 \to \kappa_{l.r.}$ as should be obvious from \eqref{eqn:FphiH} and \eqref{eqn:FrH}. In the limit $\kappa_0 \rightarrow 0$, we also recover the NHEK results \eqref{N1}-\eqref{N2}. This is most easily obtained after noticing the intermediate $\kappa_0 \rightarrow 0$ limits, 
\bea
\frac{\omega}{\kappa} &\rightarrow & -\frac{3}{4} \frac{m \hat x_0}{ \lambda}, \\
\frac{\Gamma(h-i n + i m)}{\Gamma (1-h-i n + i m) } &\rightarrow & \left( \frac{3 i m \hat x_0}{4} \lambda^{-1}\right)^{2h-1}, \\
(2\kappa)^{h} \frac{ \cR^{\text{in}}}{ \mathcal W_\kappa} & \rightarrow & (-2 i {\Omega})^{h}\frac{ \cW^{\text{in}}}{ \mathcal W},\\
{(2\kappa)^{-h} \cR^{\text{D}}} & {\rightarrow} &{(-2 i \Omega)^{-h}\cM^{\text{D}}}.
\eea
Here, the NHEK and near-NHEK Wronskian (with the convention that the first solution is the ingoing one and the second the Dirichlet one) are
\bea
\mathcal W = \frac{2 i {{\Omega}} \Gamma(2h)}{\Gamma(h-i m)}, \qquad \mathcal W_\kappa = -\frac{2\kappa \Gamma(2h)\Gamma(1-i n )}{\Gamma(h+i(m-n))\Gamma(h-i m )}.
\eea

For the inhomogenous part, a direct application of \eqref{Lo} yields 
\bea
\tilde F_{\phi}^I( \kappa_0) &=&    \sum_{\hat l, m \neq 0} \Big ( im \sqrt{3+6\kappa_0-\kappa^2_0} \cR^{\text{D}}_0 \cR^{\text{in}}_0  [S_{\hat lm}(\frac{\pi}{2})]^2 \frac{\Gamma(h-i m) \Gamma(h-i (n-m))}{4\kappa_0 \Gamma(2h) \Gamma(1-i n)} \Big), \\
\tilde F_r^{I,+}( \kappa_0) &=&  \sum_{\hat l, m} \Big ( \sqrt{3+6\kappa_0-\kappa^2_0} \cR'^{\text{D}}_0 \cR^{\text{in}}_0  [S_{\hat lm}(\frac{\pi}{2})]^2 \frac{\Gamma(h-i m) \Gamma(h-i (n-m))}{4\kappa_0^2 \Gamma(2h) \Gamma(1-i n)} \Big), \\
\tilde F_r^{I,-}( \kappa_0) &=&  \sum_{\hat l, m} \Big ( \sqrt{3+6\kappa_0-\kappa^2_0} \cR^{\text{D}}_0 \cR'^{\text{in}}_0  [S_{\hat lm}(\frac{\pi}{2})]^2 \frac{\Gamma(h-i m) \Gamma(h-i (n-m))}{4\kappa_0^2 \Gamma(2h) \Gamma(1-i n)} \Big),
\eea
where $\cR'^{\text{D}}{}_0(\kappa_0) \equiv \frac{\d \cR^{\text{D}}}{\d (r/\kappa)}(r_0)$. We recover the NHEK limit \eqref{NHEKFI} when $\kappa_0 \rightarrow 0$.

\begin{figure}[!hbt]
	\centering
	\subfigure{
		\includegraphics[width=.45\textwidth]{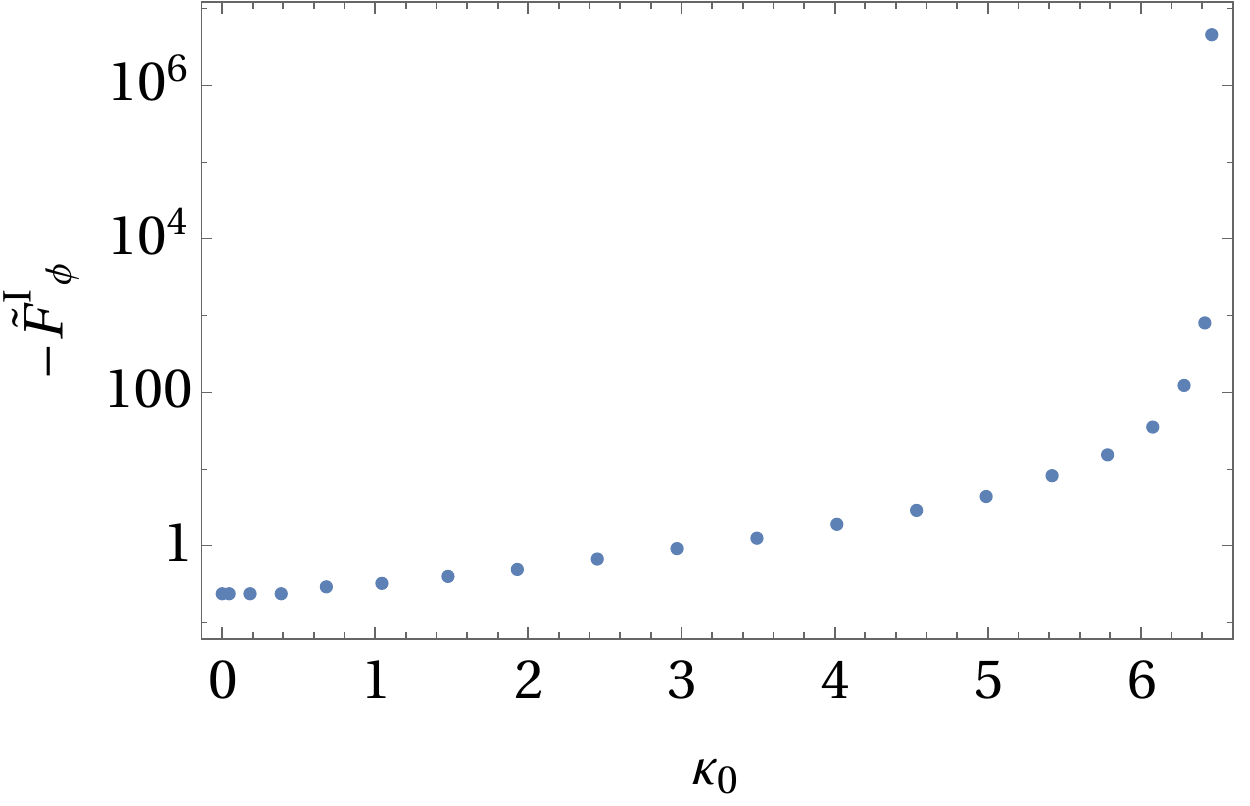}}
	\qquad
	\subfigure{
		\includegraphics[width=.45\textwidth]{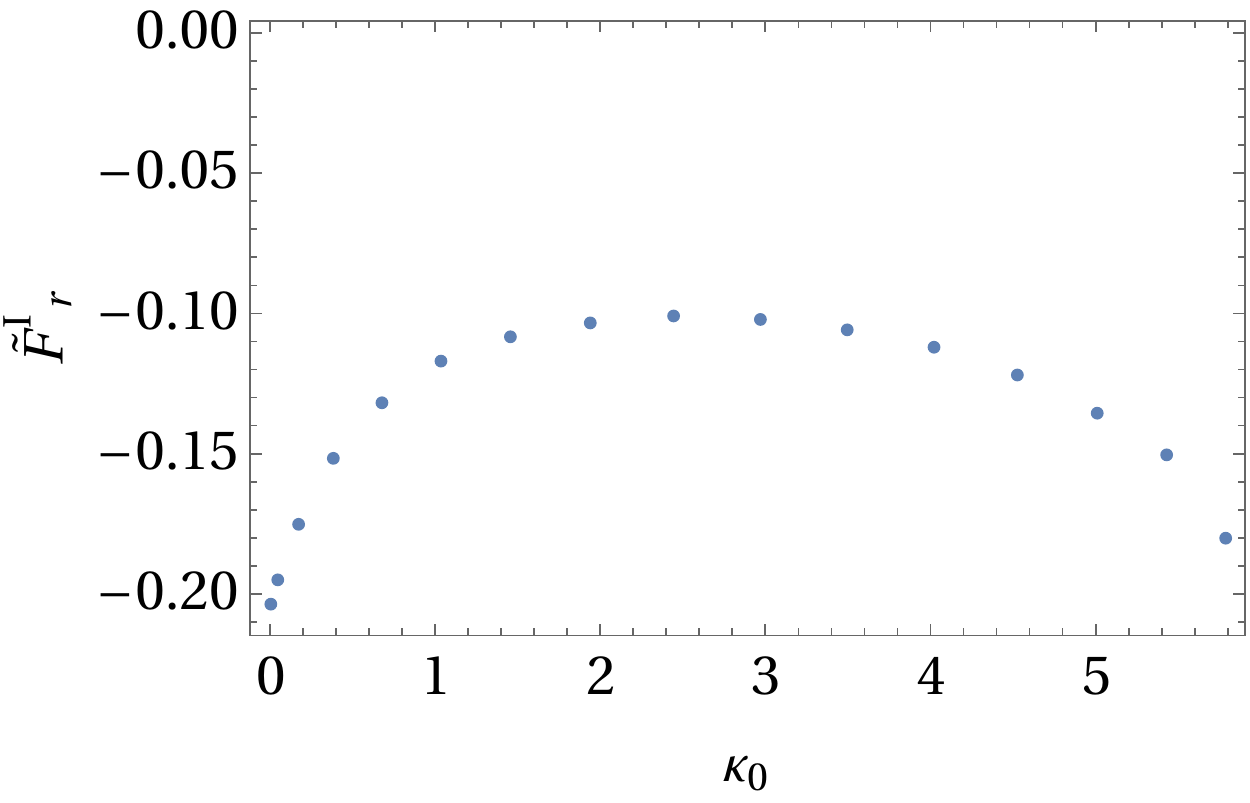}
		}
	\qquad
	\caption{$\tilde F_{\phi}^I{}(\kappa_0)$ (left) and $\tilde F_r^I{}(\kappa_0)$ (right)}
	\label{fig:FinhomnearNHEK}
\end{figure}

 The inhomogeneous piece is now not simply a number as for NHEK \eqref{FPhiI} but rather a function of $\kappa_0$. Exactly as in the NHEK case, $\tilde F_r^{I}( \kappa_0)$ requires regularization but $\tilde F_{\phi}^{I}( \kappa_0)$ does not. The regularization is discussed in Appendix \ref{app:reg}. It is again a nontrivial check that the formulas derived there exactly match the divergent pieces.  Fig. \ref{fig:FinhomnearNHEK} illustrates for a selected number of numerical points how these inhomogeneous parts behave as a function of $\kappa_0$. Especially for high $\kappa_0$, determining these points becomes numerically demanding. For the radial component of the self-force the convergence is slow as it is governed by a power law rather than an exponential. This problem also arises for the more general self-force problem in a Kerr black hole which is why there is an important effort in deriving higher order regularization coefficients even if the presently known coefficients are enough to have a converging series \cite{heffernan2014high}. In addition, here the feature arises that, closer to the light ring, more $l$ modes contribute significantly to the result. This further complicates the numerical calculation and has prevented us from finding accurate values of $\tilde F_r^I{}(\kappa_0)$ closer to $\kappa_{l.r.}$.  The numerical values for these points are given in Appendix \ref{app:num}, specifically on Table \ref{tbl:nearFtildephi} and Table \ref{tbl:nearFtilder}. Similarly to what was done for the NHEK results, we also give an estimate of the truncation error $\eps_{trunc}$ by computing $\tilde{F}^{(\text{tail})}_{\phi} \approx \sum^{\infty}_{\hat l= \hat l_{\text{cutoff}}+1} a_0 e^{-a_1 \hat l}$, $\tilde{F}^{(\text{tail})}_{r} \approx \sum^{\infty}_{ l=l_{\text{cutoff}}+1} \frac{a_2}{l^2}$ where the coefficients $a_0, a_1, a_2$ are determined from a fit to the high $\hat l$ (respectively $l$) data points. As a consistency check, note that for $\kappa_0=0$, the NHEK limit is recovered. Furthermore, all (unstable) circular orbits experience an attractive radial self-force, in line with the observation of \cite{warburton2010self} that this is also true for stable circular geodesics when $a \geq 0.461 M$.

\section{Self-force for equatorial orbits}
\label{sec:SF}

With the self-force for sources on circular orbits available, we now turn toward generic timelike equatorial orbits incoming from the asymptotically flat region. As shown in \cite{Compere:2017hsi}, these can be connected to the circular orbits by conformal transformations. Given the self-force in a seed spacetime $F^{\mu}_{\text{self,seed}}$, with coordinates $X^{\mu}$, in which the orbit is circular and a diffeomorphism $X^{\mu}(\bar x^{\alpha})$ mapping this orbit to a more general geodesic in the near-horizon region, the self-force has been conjectured to be given at leading order in the high spin limit by the tensorial transformation law induced by the coordinate transformation  \cite{Hadar:2016vmk}
\be
F^{\text{self,physical}}_{\alpha}(\bar x^\beta) =  F^{\text{self,seed}}_{\mu} \frac{\partial X^{\mu}}{\partial \bar x^{\alpha}}.
\label{eqn:simpleSF}
\ee
Now, it turns out that this is not exactly the case. Since the traveling waves that couple the near-horizon region with the asymptotic region are not conformally covariant, they are transformed nontrivially under the coordinate transformation. A correction to the homogeneous part of the perturbation is therefore required in order to restore the appropriate boundary conditions that correspond to no ingoing radiation in the asymptotically flat region.  Since the homogeneous part of the perturbation is regular, it does not require a regularization and it can be obtained directly from the boundary conditions for the final orbit. In contrast, a solution with Dirichlet boundary conditions transforms covariantly under conformal transformations. We therefore propose the following scheme. We can use \eqref{eqn:simpleSF} to obtain the self-force for an arbitrary equatorial orbit with Dirichlet boundary conditions by starting with the seed circular solution with Dirichlet boundary conditions. All such conformal maps were detailed in  \cite{Compere:2017hsi}. We will then add to the resulting self-force the self-force caused by the homogeneous correcting solution to the linear field equations that is required in order to respect the boundary conditions. In this section, we will make the foregoing strategy explicit and give an example of its application.

Luckily, as in the circular case, the self-force caused by the homogeneous correcting solution is numerically negligeable (it bring corrections $ < 10^{-4}$), as we will illustrate. Therefore, up to small but nonvanishing corrections, the formula \eqref{eqn:simpleSF} will still be valid.

\subsection{Summary of self-force data on circular orbits}
\label{sec:sumcirc}
 The first step in the outlined procedure above relies on the self-force data on circular orbits. We obtained all such data in the previous section. For convenience, we summarize the main results here. The NHEK and near-NHEK radial self-force are given by
\bea
F_R &=& \frac{q^2}{M R_0} \tilde{F}_R,  \qquad \qquad F_r =  \frac{q^2}{M r_0} \tilde{F}_r, \nn \\
\tilde{F}_R &\approx& \tilde{F}^I_R \approx -0.204 ,\qquad   \tilde{F}_r \approx \tilde{F}^I_r(\kappa_0) , \quad \text{(see Fig. \ref{fig:FinhomnearNHEK} and Table \ref{tbl:nearFtilder})}  .
\eea
The NHEK and near-NHEK angular self-force are given by	
\bea
F_{\Phi} &=& \frac{q^2}{M} \tilde{F}_{\Phi} ,\qquad \qquad \;\;\;\;F_{\phi} =  \frac{q^2}{M} \tilde{F}_{\phi},\nn \\
\tilde{F}_{\Phi} &\approx& \tilde{F}^I_{\Phi} \approx -0.225 ,\qquad  \tilde{F}_{\phi} \approx  \tilde{F}_{\phi}^I(\kappa_0) ,\quad \text{(see Fig. \ref{fig:FinhomnearNHEK} and Table \ref{tbl:nearFtildephi})} .
\eea
The time and polar components are derived from 
\be
F_T = -\tilde{\Omega} F_{\Phi} , \qquad F_t = -\tilde{\omega} F_{\phi},\qquad F_\theta = 0.
\ee
where the angular frequencies are 	
\be
\tilde{\Omega} = -\frac{3 R_0}{4} , \qquad \tilde{\omega} = -\frac{3 r_0}{4}(1+\kappa_0).
\ee

\subsection{Dirichlet self-force}
	
    The inhomogeneous contribution to the self-force with Dirichlet boundary conditions as defined in Sec. \ref{sec:circular} is insensitive to the boundary in the asymptotically flat space and can be transformed straightforwardly under the symmetries of the near-horizon region such that we can simply apply \eqref{eqn:simpleSF}, i.e.
    
	\be
	F_{\alpha}^{\text{self,physical}, I} =  F_{\mu}^{\text{self,seed}, I} \frac{\partial X^{\mu}}{\partial \bar x^{\alpha}}.
	\label{eqn:inhomsimpleSF}
	\ee
    
     Let us briefly review how the symmetries relate different geodesics in (near-)NHEK with circular orbits. An equatorial geodesic in (near-)NHEK is characterized by its energy $E$ ($e$), with respect to the (near-)NHEK Killing vector $\partial_T$ ($\partial_t$), and its specific angular momentum $\ell$. There are two types of equivalence classes under the near-horizon $SL(2,\mathbb{C}) \times U(1) \times PT$ symmetry for incoming geodesics from the exterior Kerr region. The first contains critical orbits with $\ell=\ell^*$ and therefore in particular the circular NHEK orbit \cite{Compere:2017hsi}:
    
    \begin{itemize}
    	\item[$ $] Circular$_*$ (ISCO) $\Leftrightarrow$ Plunging$_*(E)$  $\Leftrightarrow$ Plunging$_*(e=0)$ $\Leftrightarrow$ Plunging$_*(e)$
    \end{itemize}
    
The second contains more general supercritical orbits $\ell > \ell^*$ and in particular the near-NHEK circular orbits 
    
    \begin{itemize}
    	\item[$ $] Circular$(\ell)$ $\Leftrightarrow$ Marginal$(\ell)$ $\Leftrightarrow$ Osculating$(E,\ell)$ $\Leftrightarrow$  Plunging$(E,\ell)$ $\Leftrightarrow$ Osculating$(e,\ell)$ $\Leftrightarrow$  Plunging$(e,\ell)$ 
    \end{itemize}
    
	We will refer to \cite{Compere:2017hsi} (in particular Appendix B) for the explicit description of all the equatorial orbits and their relations and to \cite{Compere:2019bb} for a description of nonequatorial orbits and subcritical orbits $\ell < \ell_*$. Later, in Sec. \ref{sec:example}, we will treat the Plunging$_*(e=0)$ orbit in detail. In the remainder of this section, we will instead apply \eqref{eqn:inhomsimpleSF} specifically to the Plunging$(e,\ell)$ trajectories. These are a family of near-NHEK geodesics parametrized by $e > -\frac{\sqrt{3} \kappa}{2} \sqrt{\ell^2-\ell^2_*}$ and $\ell > \ell_*$. Other orbits could of course be treated similarly but should not give rise to any further qualitatively distinct behavior. The Plunging$(e,\ell)$ geodesics can be described by
	
	\bea
	\bar \tau &=& \bar \tau_0-\frac{1}{2} \log (1+\frac{2}{\bar \rho}) +\frac{1}{2} \log \frac{(3 \ell^2-4 M^2-4 \eps \ell )\bar \rho+2( \ell^2 -2 \eps^2 -4 M^2+(\eps- \ell)F)}{(3	\ell^2-4 M^2+4 \eps \ell )\bar \rho +4(\eps+ \ell)^2 - 2 (\eps+ \ell)F},\nn\\
	\bar \phi &=&\bar \phi_0-\frac{1}{2}\log(\bar \rho(\bar \rho+2))+\frac{\sqrt{3}\ell}{2\sqrt{\ell^2-\ell_*^2}}\log(3(\ell^2-\ell^2_*)(\bar \rho+1)+\sqrt{3(\ell^2-\ell^2_*)}F+4 \eps \ell) \nn\\
	&& +\frac{1}{2}\log ( (7\ell^2-4M^2)\bar \rho (\bar \rho+2)+16 \eps \ell +8 (\eps+\ell )^2 -4(\eps+\ell(\bar \rho+1))F),\label{gen7}
	\eea
	where $F=\sqrt{3(\ell^2-\ell_*^2)\bar \rho (\bar \rho +2)+8 \eps \ell \bar \rho+ 4 (\eps+ \ell)^2}$. Here, we have used the rescaled quantities $\bar \tau = \kappa \bar t$, $\bar \rho = \bar r /\kappa$, $\eps = e/\kappa$ to eliminate $\kappa$. The transformation relating it to a circular orbit \eqref{nearNHEKcirc} was first given by \cite{Hadar:2016vmk}
	\bea
	\frac{r}{\kappa} &=& \sqrt{\bar \rho(\bar \rho+2)} (\sinh  \bar \tau + \chi \cosh \bar \tau) - \chi (\bar \rho + 1) - 1 ,\nn \\
	\kappa t &=&  \log \frac{\sqrt{\bar \rho(\bar \rho+2)} \cosh \bar \tau - (\bar \rho + 1)}{\sqrt{\rho ( \rho + 2)}}\label{eq:1}, \\
	\phi &=& \bar \phi - \frac{1}{2}\log\left(  \frac{\sqrt{\bar \rho(\bar \rho+2)} - (\bar \rho + 1) \cosh \bar \tau + \sinh \bar \tau}{\sqrt{\bar \rho(\bar \rho+2)} - (\bar \rho + 1) \cosh  \bar \tau -  \sinh \bar \tau} \frac{  \rho +2}{  \rho}\right), \nn
	\eea
	 with
	\bea
	\eps &=&  \frac{1}{2}\sqrt{3(\ell^2-\ell_*^2)} \chi,\qquad  \ell>\ell_*, \label{defechi}\\
	\bar{\tau}_0&\equiv & \kappa \bar t_0 = -\frac{1}{2}\log \frac{1+\chi}{1-\chi}. 
	\eea
	
	 Here, the parameter $\chi$, which could be interchanged for $e$, must satisfy $\chi>-1$ in order for the orbit to be plunging but for $|\chi| > 1$, $\bar t_0$ is complex, and therefore a complex shift of $\bar t \rightarrow \bar t \pm i \pi$ is required as a final step. It is straightforward to write out \eqref{eqn:inhomsimpleSF} explicitly but since it becomes rather elaborate, we shall instead highlight the behavior of the self-force in various limits and specific cases. 
	 
 An important fact to note at the outset is that the angular component $F_{\bar \phi}$, and therefore the angular momentum flux, is the same as that for the circular orbit. This turns out to be true for all orbits described in \cite{Compere:2017hsi} as a consequence of the explicit form of the conformal transformations. We therefore proved that the angular momentum flux is universal at leading order in the high spin limit for all equatorial orbits.
	 
\begin{figure}[!hbt]
	 	\centering
	 	\subfigure{
	 		\includegraphics[width=.45\textwidth]{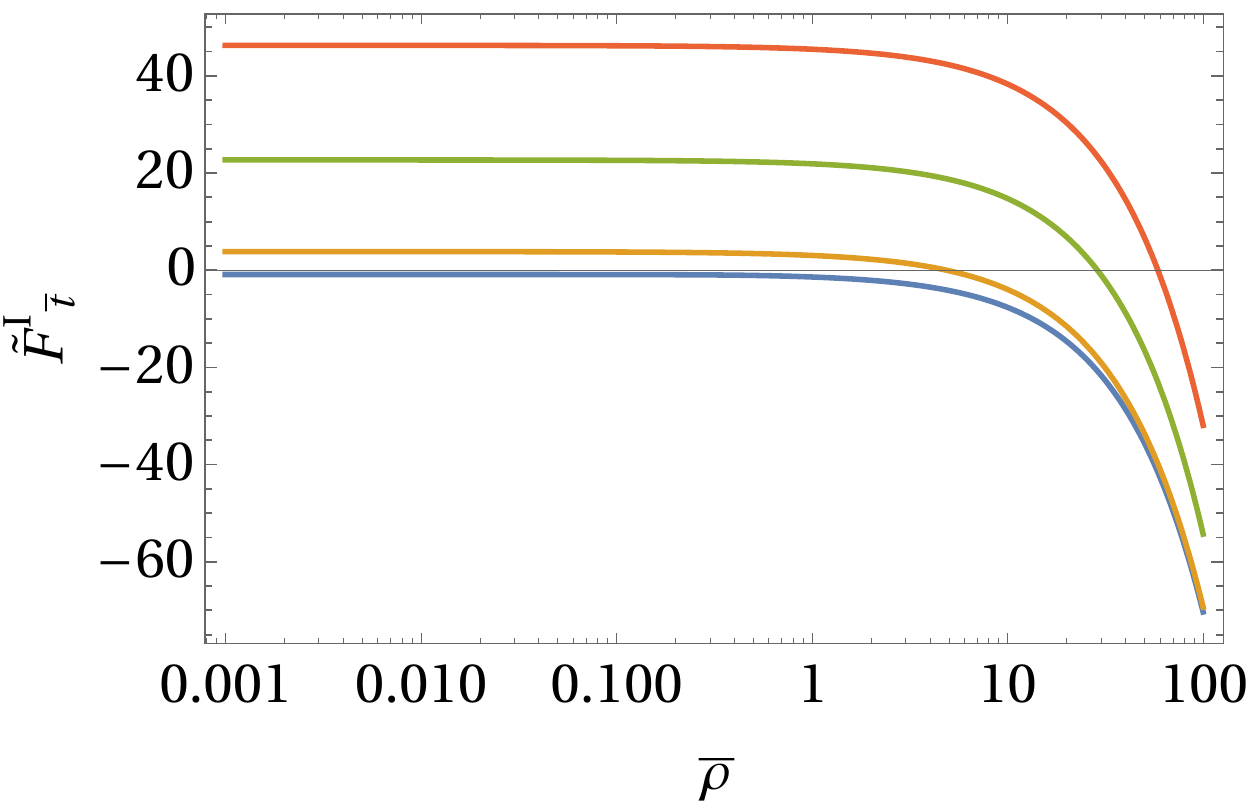}}
	 	\qquad
	 	\subfigure{
	 		\includegraphics[width=.45\textwidth]{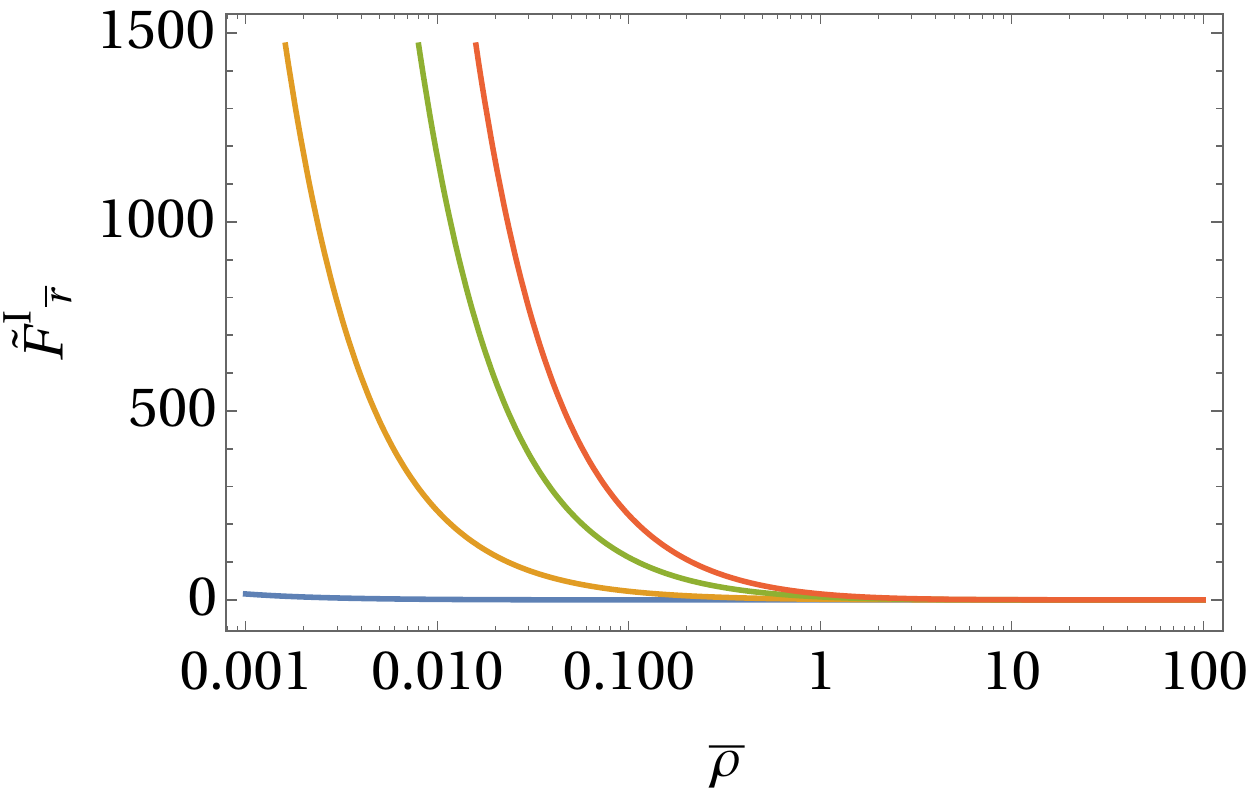}}
	 	\qquad
	 	\caption{$\tilde{F}^I_{\bar t}\equiv \frac{M}{q^2 \kappa}F^I_{\bar t}$ (left) and $\tilde{F}^I_{\bar r}\equiv\frac{M \kappa}{q^2}F^I_{\bar r}$ (right)  along a Plunging$(e,2\ell_*)$ orbit with $e \kappa^{-1}$ given by $\eps_*+0.1M$ (lowest, blue), $\eps_*+100M$ (second lowest, orange), $\eps_*+500M$ (second uppermost, green) and $\eps_*+1000M$ (uppermost, red), where $\eps_* \equiv -\frac{\sqrt{3}\ell_*}{2}$. }
	 	\label{fig:Finhomplunging}
	 \end{figure}

	 Consider now for instance $\ell = 2\ell_*$ for various values $e$. The nontrivial components of the self-force $\tilde{F}^I_{\bar r}$, $\tilde{F}^I_{\bar t}$ are given in Fig. \ref{fig:Finhomplunging}.
	 
For fixed $e$, $\ell$ it is straightforward to understand the behavior as $\bar \rho \to 0$ at fixed $\bar \tau$ 
	 
	 \bea
	 F^I_{\bar t}(\bar \rho \to 0) &\sim&\frac{q^2 \kappa}{M} (\frac{3\ell^2_*-\eps \ell }{3\ell^2_*+\ell^2}\tilde{F}^I_{\phi} + \frac{2(\eps +\ell)\kappa_0}{\sqrt{3(\ell^2-\ell_*^2)}}\tilde{F}^I_r), \label{eqn:plungingneart}\\
	 F^I_{\bar r}(\bar \rho \to 0) &\sim& \frac{q^2 }{M \bar \rho \kappa} \frac{\eps+\ell}{\sqrt{3(\ell^2-\ell_*^2)}} (\kappa_0 \tilde{F}^I_r-\frac{\ell \sqrt{3(\ell^2-\ell_*^2)}}{2 (3\ell^2_*+\ell^2)}\tilde{F}_{\phi}). \label{eqn:plungingnearr}
	 \eea
	 On the other hand, as $\bar \rho \to \infty$ 
	 \bea
	 F^I_{\bar t}(\bar \rho \to \infty) &\sim& \frac{q^2 \bar \rho \kappa}{M}  ( \frac{3 \ell^2_*}{3\ell_*^2+\ell^2} \tilde{F}^I_{\phi}+ \kappa_0 \tilde{F}^I_r ) ,\label{eqn:plungingfart} \\
	 F^I_{\bar r}(\bar \rho \to \infty) &\sim& \frac{q^2 }{M \bar \rho \kappa} \frac{2 \ell}{\sqrt{3(\ell^2-\ell_*^2)}} ( \kappa_0 \tilde{F}^I_r - \frac{3(\ell^2-\ell^2_*)}{4 (3\ell^2_*+\ell^2)} \tilde{F}^I_{\phi}). \label{eqn:plungingfarr}
	 \eea

\begin{figure}[!hbt]
	\centering
	\subfigure{
		\includegraphics[width=.45\textwidth]{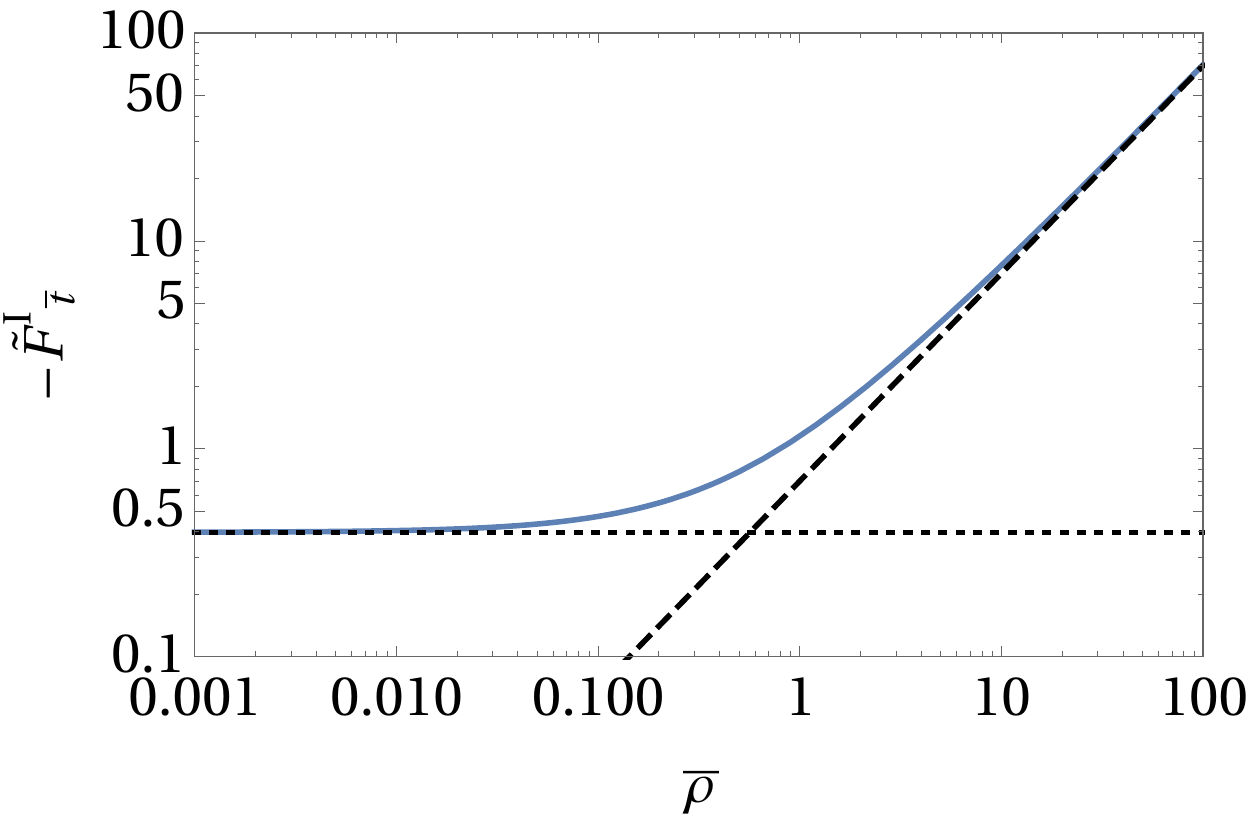}}
	\qquad
	\subfigure{
		\includegraphics[width=.45\textwidth]{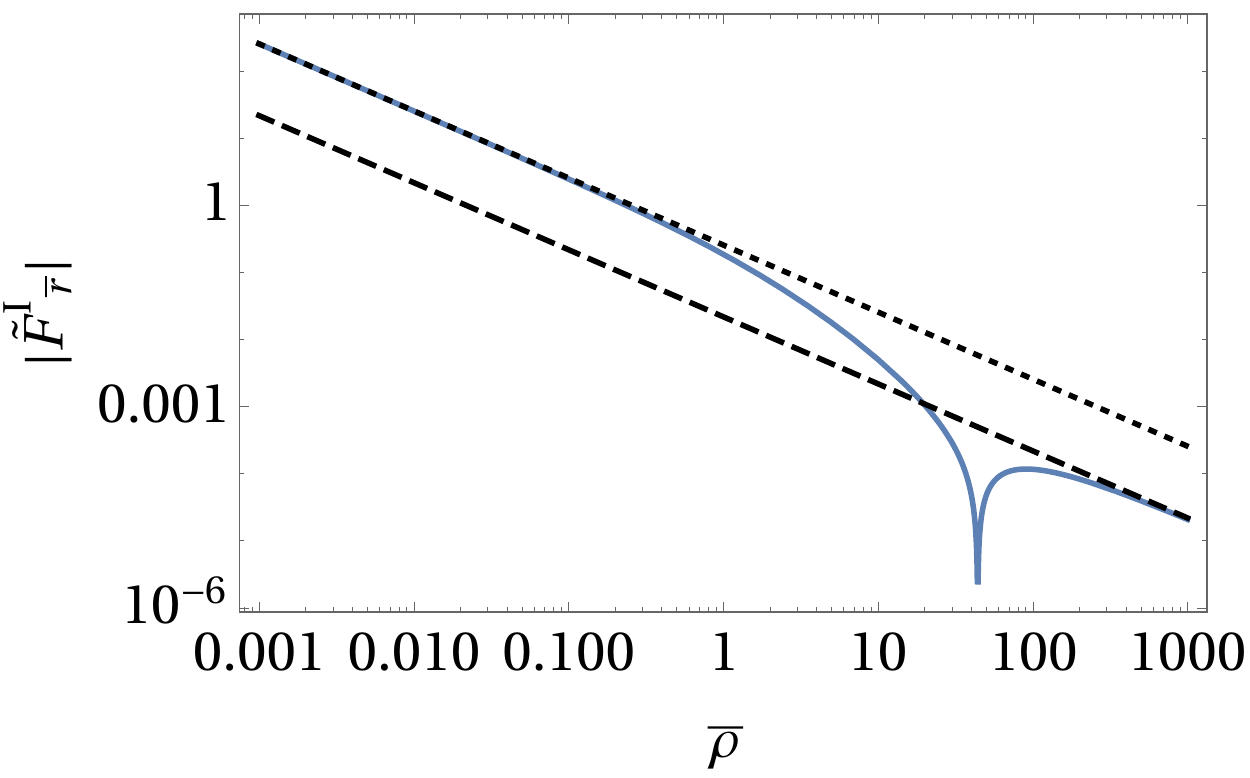}}
	\qquad
	\caption{$\tilde{F}^I_{\bar t}\equiv \frac{M}{q^2 \kappa}F^I_{\bar t}$ (left) and $\tilde{F}^I_{\bar r}\equiv \frac{M \kappa}{q^2}F^I_{\bar r}$ (right)  along a Plunging$(\kappa\frac{5\sqrt{3}\ell_*}{2},2\ell_*)$ orbit (solid blue) as compared to the asymptotic approximations \eqref{eqn:plungingfart},\eqref{eqn:plungingfarr} (dashed) and \eqref{eqn:plungingneart},\eqref{eqn:plungingnearr}  (dotted).}
	\label{fig:Finhomplungingasymptotics}
\end{figure}

	 This latter asymptotic behavior is independent of $e$ but its range of validity is not, as can be confirmed by looking at Figure \ref{fig:Finhomplunging}. As $\ell \to \ell_*$ the self-force becomes asymptotically identical to that of a NHEK circular orbit after taking	
	  $\bar \rho \sim \frac{1}{\ell^2-\ell_*^2}$ as required to ensure the applicability of the asymptotic expansion \eqref{eqn:plungingfart},\eqref{eqn:plungingfarr}. From Fig. \ref{fig:Finhomplungingasymptotics}, it can be seen that the actual complete self-force does little more than interpolate between the asymptotic behaviors \eqref{eqn:plungingneart},\eqref{eqn:plungingnearr} and \eqref{eqn:plungingfart},\eqref{eqn:plungingfarr}.

	  As far as we are aware, this level of control of the self-force along a noncircular trajectory around a black hole can perhaps only be matched by post-Newtonian post-Minkowskian expansions \cite{hopper2016analytic, bini2016new}. However, as has been pointed out, it is not the full story. Indeed, given the absence of a complete decoupling of the asymptotically flat region, even to leading order in $\lambda$, this self-force needs a correction, as we will discuss next. 
	 	
	\subsection{Correcting homogeneous self-force}
	\label{sec:genhom}
		
	 The final step in constructing the self-force is ensuring that the appropriate outgoing boundary conditions in the asymptotically flat spacetime are satisfied for the final solution. This amounts to fixing a mixed Dirichlet-Neumann boundary condition for each mode $\hat l,m,\Omega$. The amplitude $A_{\hat lm\Omega}(\gamma) $ of the ingoing mode that needs to be added accordingly as a function of the asymptotic ($\bar R > \bar R_0 $) Dirichlet mode $B_{\hat lm\Omega}(\gamma)$, which is defined for NHEK at large radius $R$ for a given worldline $\gamma$ through
	
	\be
	\Psi \sim \frac{1}{\sqrt{2 \pi}} \int^{\infty}_{-\infty} \d \Omega \sum_{\hat lm} (A_{\hat lm\Omega} (\gamma) \cW^{\text{in}}_{\hat lm\Omega}(R)+B_{\hat lm\Omega} (\gamma)\cM^{\text{D}}_{lm\Omega}(R) ) S_{\hat l m}(\theta) e^{i m \Phi-i\Omega T},
	\ee
	is given by \cite{Compere:2017hsi}\footnote{This is in particular how $Y$ was found in \eqref{eqn:solR} and analogously in the near-NHEK circular solution \eqref{eqn:nearRadialSolution}.}
	\bea
	A_{\hat lm\Omega}(\gamma) =Y(\lambda^{2/3}\Omega) B_{\hat lm\Omega}(\gamma)
	\label{eqn:A}
\eea
where $Y$ was defined in terms of $\lambda^{2/3}\Omega$ in \eqref{eqn:Y}. Note that $\Omega$ and $B_{\hat lm\Omega}(\gamma)$ should not be replaced here by their circular values. 
	
	The homogeneous piece of the self-force that still needs to be added to the inhomogeneous piece described previously is then
	\be
	F^H_{\bar \mu} = \frac{q}{\sqrt{2 \pi}}  \partial_{\bar \mu} (\int^{\infty}_{-\infty} \d \Omega \sum_{\hat lm} A_{\hat lm\Omega} (\gamma) \cW^{\text{in}}_{\hat lm\Omega}(\bar R)S_{\hat l m}(\theta) e^{i m \bar \Phi-i\Omega \bar T}).
	\label{eqn:genhom}
	\ee 
	
Similarly, in the case of near-NHEK at large radius $r$, the complete linear solution is
	\be
	\Psi = \frac{1}{\sqrt{2 \pi}} \int^{\infty}_{-\infty} \d \omega \sum_{\hat lm} (A_{\hat lm\omega} (\gamma)  \cR^{\text{in}}_{\hat lm\omega}(r)+B_{\hat lm\omega} (\gamma) \cR^{\text{D}}_{\hat lm\omega}(r)) S_{\hat l m}(\theta) e^{i m \phi-i\omega t},
	\ee
	with
	\be
	A_{\hat lm\omega} (\gamma)= Y(\lambda,\omega ) B_{\hat lm\omega}(\gamma) \label{eqn:nearA}
	\ee
where $Y(\lambda,\omega )$ is defined in \eqref{Ynear}. Therefore, the correcting homogeneous self-force to the Dirichlet self-force is given by 
	\be
	F^H_{\bar \mu} = \frac{q}{\sqrt{2 \pi}}  \partial_{\bar \mu} (\int^{\infty}_{-\infty} \d \omega \sum_{\hat lm} (A_{\hat lm\omega} (\gamma) \cR^{\text{in}}_{\hat lm\omega}(\bar r)S_{\hat l m}(\theta) e^{i m \bar \phi-i\omega \bar t}).
	\label{eqn:gennearhom}
	\ee 
		
	In Appendix \ref{app:Bs}, we give the expressions for $B_{\hat lm\omega}(\gamma)$ and $B_{\hat lm\Omega}(\gamma)$ for each family of equatorial geodesics. These expressions can almost immediately be read of from the analogous gravitational cases given in \cite{Compere:2017hsi}. Remark first that from \eqref{eqn:genhom}, \eqref{eqn:gennearhom} one can conclude  from the expression for $Y$ \eqref{eqn:Y} that the contribution due to the normal modes (which have $h >\frac{1}{2}$) are indeed parametrically subleading. The crucial observation, also stressed in \cite{Compere:2017hsi}, is that consistency of the near-horizon approach requires that the integrals are effectively cut-off at $\Omega \sim \pm \lambda^{-2/3}$ such that $\lambda^{2/3} \Omega$ can always be considered small in NHEK. Similarly, in near-NHEK, there should not be a significant contribution from $\omega \sim \pm \lambda^{-1}$ such that $\lambda \omega$ can be considered small in the entire integration range, which implies also that $Y$ \eqref{Ynear} is subleading as $\lambda \rightarrow 0$ for normal modes with respect to traveling waves.
	
	In the near-NHEK case, we can approximate the integral \eqref{eqn:gennearhom} at late times as a sum over quasinormal modes (QNMs), which correspond to simple poles in $Y$. They are approximately given by (see also \cite{Yang:2013uba})
	\be
	\omega_{Nlm} = -i\kappa(N+h) + i\kappa (-1)^N \frac{\lambda^{2h-1} k_2}{N! \Gamma(1-2h-N)}.
	\label{eqn:QNM}
	\ee
	We have included the first order correction to stress the difference between the two types of modes: the normal modes ($h > 1/2$) for which the correction becomes parameterically small,  and the traveling waves ($\text{Re}(h)=1/2$) for which it does not. Remark moreover that $B_{\hat lm\omega}$ contains an additional factor $\Gamma(h-in+im)=\Gamma(h- \frac{i \omega}{\kappa})$ leading to poles at $\omega_{Nlm} =-i\kappa(N+h)$. This means that for the traveling wave modes one actually has two close lying first order poles, even as $\lambda \to 0$, as opposed to second order poles. The observed pole structure is in fact characteristic of the type of separation we have performed. By considering first the unphysical Dirichlet boundary conditions, we have introduced spurious poles which are associated to the vanishing of the Wronskian between $\cR_{\hat lm \omega}^{D}$ and $\cR_{\hat lm \omega}^{\text{in}}$. The additional piece in the solution then has two types of poles, associated on the one hand to the set of physical QNMs and on the other hand a set whose sole function is to cancel the spurious poles. 
	
\subsection{Master formula}

Assembling the two building blocks described previously, the self-force on generic equatorial timelike orbits with a seed orbit in NHEK at leading order in high spin is given by our master formula
\be
F_{ \mu}(\bar{x}^{\mu}) = F^{I}_{\nu}J^{\nu}{}_{\mu}(\bar{x}^{{\mu}})  + \frac{q}{\sqrt{2 \pi}} \int \d \Omega \sum_{lm} A_{lm\Omega} \partial_{\mu} (\cW^{\text{in}}_{lm\Omega}(\bar{R}) S_{l m}(\theta) e^{i m \bar{\Phi}-i\Omega \bar{T}} )
\label{eqn:masterNHEK}
\ee
with reference to Sec. \ref{sec:sumcirc}, Appendix \ref{app:Bs}, and the Appendix B of \cite{Compere:2017hsi} to find respectively the needed input $F^{I}_{\nu}$, $A_{lm\Omega}(\gamma)$ depending on the desired geodesic $\gamma$ and $J^{\nu}{}_{\mu}= \p x^\nu / \p \bar x^\mu$. Here barred coordinates $\bar{x}^{\mu}$ denote coordinates in the final target NHEK spacetime where the orbits lie. 

The analogous master formula for a final orbit in near-NHEK reads as
\be
F_{ \mu}(\bar{x}^{\mu}) = F^{I}_{\nu}J^{\nu}{}_{ \mu}(\bar{x}^{{\mu}})  + \frac{q}{\sqrt{2 \pi}} \int \d \omega \sum_{lm} A_{lm\omega} \partial_{\mu} (\cR^{\text{in}}_{lm\omega}(\bar{r}) S_{l m}(\theta) e^{i m \bar{\phi}-i\omega \bar{t}} ).
\label{eqn:masternearNHEK}
\ee

\subsection{An example}
\label{sec:example}

As a fully worked out example we will consider a Plunging$_*(e=0)$ geodesic. This is a trajectory in near-NHEK described by

\bea
\bar t &=& \bar t_0-\frac{1}{2\kappa}\log \left( \bar r(\bar r+2\kappa) \right),\\
\bar \phi &=& \bar \phi_0 +\frac{3}{4\kappa} \bar r -\frac{1}{2}\log (1+\frac{2\kappa}{\bar r}).\label{Plungingstare0}
\eea

 It is related to the NHEK circular orbit by
\bea
T &=& -e^{-\kappa  \bar t} \frac{ \bar r + \kappa}{\sqrt{ \bar r ( \bar r+2\kappa)}},\nn \\
R &=& \frac{1}{\kappa} e^{\kappa  \bar t} \sqrt{ \bar r ( \bar r + 2\kappa)},\label{NHEKtonearNHEK} \\
\Phi &=&  \bar \phi - \frac{1}{2} \log \frac{ \bar r }{ \bar r +2 \kappa}. \nn
\eea

The parameters of the orbits are related by
\bea
R_0=\frac{1}{\kappa} e^{\kappa \bar t_0},\qquad \Phi_0= \bar \phi_0-\frac{3}{4}.\label{mapp1}
\eea 

One finds the Jacobian
\be
J^{\nu}{}_{\mu} = \frac{\partial X^{\nu}}{\partial \bar{x}^{\mu}} = \begin{bmatrix}
	\frac{e^{- \kappa \bar t} \kappa (\kappa +  \bar r)}{\sqrt{ \bar r ( \bar r + 2 \kappa)}}	&	 \frac{e^{-\kappa \bar t} \kappa^2}{( \bar r (\bar r + 2\kappa))^{3/2}}& 0 & 0 \\
	e^{\kappa  \bar t} \sqrt{\bar r (2 \kappa + \bar r)} & \frac{e^{\kappa  \bar t} (\kappa + \bar r)}{\kappa \sqrt{ \bar r ( \bar r + 2\kappa)}} & 0 & 0 \\
	0 & 0 & 1 & 0 \\
	0 &   - \frac{\kappa}{(2 \kappa + \bar r) \bar r} & 0 & 1 
\end{bmatrix} .
\ee

The forces resulting from the inhomogeneous Dirichlet part of the self-force along a Plunging$_*(e=0)$, as a function of $\bar r$ are then given by
\bea
F^I_{\bar t}&=& \frac{q^2}{M} \Big(\frac{3}{4} (\kappa +  \bar r) \tilde F^I_{\Phi} + \kappa \tilde F^I_R \Big), \\
F^I_{\bar r}&=& \frac{q^2 (\kappa + \bar r)}{M \bar r ( \bar r + 2\kappa)} \Big(\tilde{F}^I_R - \frac{\kappa \tilde F^I_{\Phi}}{4(\bar r + \kappa)} \Big),\\
F^I_{\bar \phi}&=& \frac{q^2}{M} \tilde{F}^I_{\Phi}.
\eea

\begin{figure}[!hbt]
	\centering
\subfigure{
	\includegraphics[width=.45\textwidth]{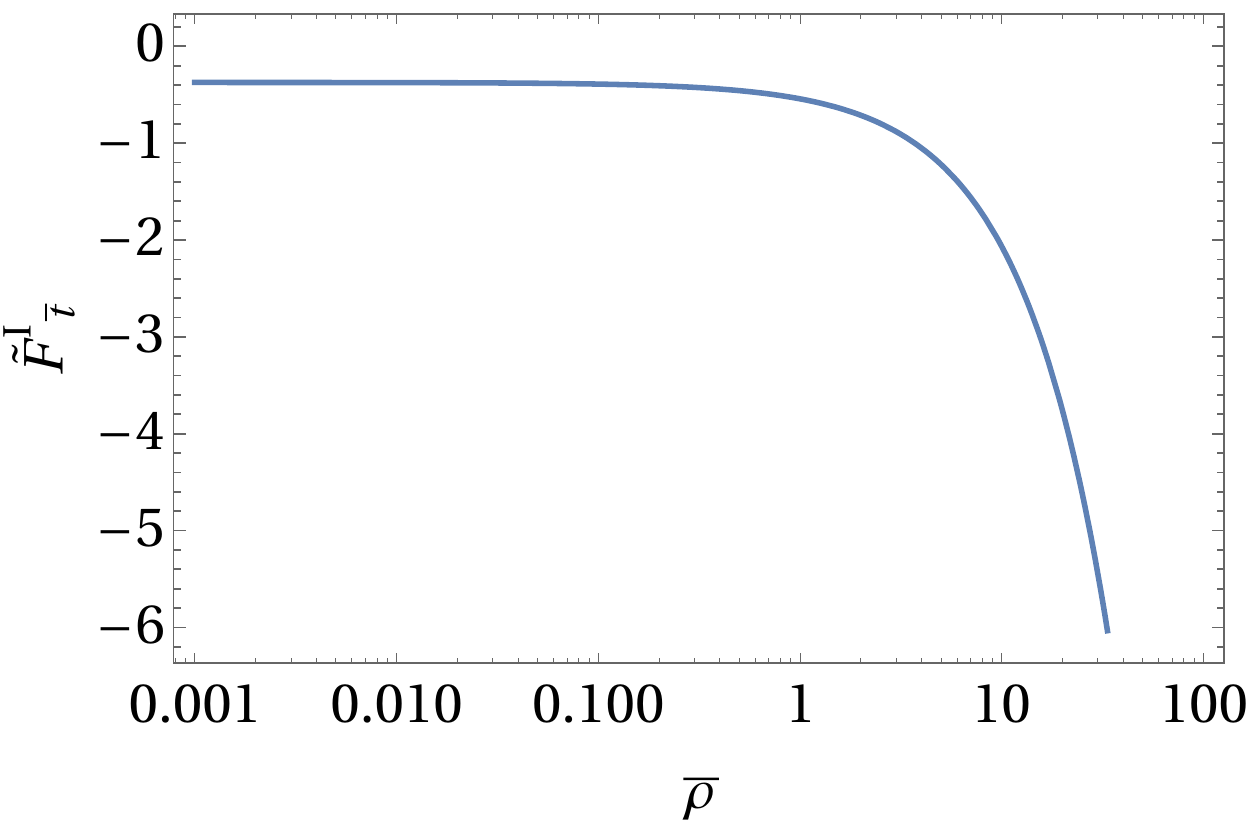}}
\qquad
\subfigure{
	\includegraphics[width=.45\textwidth]{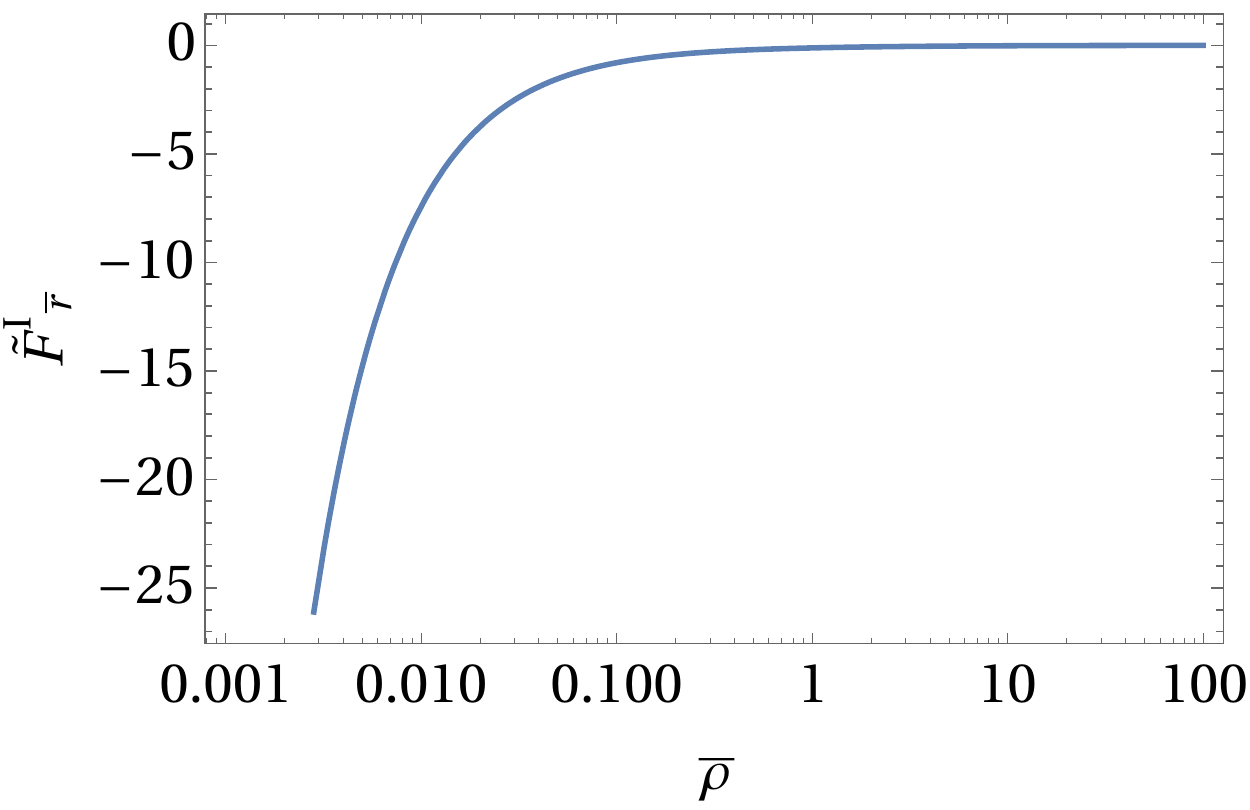}}
\qquad
	\caption{Variation along the Plunging$_*(e=0)$ trajectory of the 
		$F^I_{\bar t} \equiv \frac{q^2 \kappa}{M}\tilde{F}^I_{\bar t}$, $F^I_{\bar r} \equiv \frac{q^2}{M \kappa}\tilde{F}^I_{\bar r}$ components of the self-force with $\bar \rho = \frac{\bar r}{\kappa}$.}
	\label{fig:DForceplungee}
\end{figure}

This is illustrated in Fig. \ref{fig:DForceplungee}.

Consider now the more complicated second part of \eqref{eqn:masternearNHEK} using \eqref{eqn:nearA}, \eqref{eqn:Bplungee0}, where we focus on the traveling wave modes as the others are parametrically subleading
\be
F^H_{\bar \nu} = \frac{q}{\pi} \int \d \omega \sum_{lm} (2\kappa)^{-1} (-im\tilde{\Omega}) ^{i\omega/\kappa} B_{\text{circ}}\Gamma(h-i\frac{\omega}{\kappa}) Y(\omega,\lambda) \partial_{\bar{\nu}} (\cR^{\text{in}}_{lm\omega}(\bar{r}) S_{l m}(\theta) e^{i m \bar{\phi}-i\omega \bar{t}} ),
\ee
with
\be
B_{\text{circ}} =\frac{q S_{\hat lm}(\pi/2)}{\sqrt{3} M}  \frac{\Gamma(h-im)}{im \Gamma(2 h)}  \cW^{\text{in}}(R_0).
\ee

We will approximate the integral by a sum over the (approximately) second order poles at $\omega_N = - i \kappa (N+h)$
\be
F^H_{\bar \nu} \approx -iq \kappa \sum_{lmN} k_2 \lambda^{2h-1}\frac{ B_{\text{circ}} \Gamma(1-h-im)}{\Gamma(1-2h) N!^2}  \partial_{\omega} \big( \frac{(-im\tilde{\Omega}) ^{i\omega/\kappa}}{ \Gamma(1-in)} \partial_{\bar{\nu}} (\cR^{\text{in}}_{lm\omega}(\bar{r}) S_{l m}(\theta) e^{i m \bar{\phi}-i\omega \bar{t}} )\big)_{|\omega_N}
\ee
which is evaluated on the orbit \eqref{Plungingstare0}. 

The remaining sum does not seem to be calculable explicitly, and it is badly behaved numerically, especially as $\bar r \to \infty$. We can however, rather straightforwardly perform an asymptotic expansion. As $\bar r \to 0$, the leading contribution is given by 
\be
F^{H}_{\bar \nu}(\bar r \to 0) \sim \frac{q^2}{\sqrt{3} M} \sum_{lm} \lambda^{2h-1} k_2 (i m \frac{3}{2})^h [S_{l m}(\frac{\pi}{2})]^2  \cW^{\text{in}}_0 \frac{ \Gamma(1-h-im) \Gamma(h-im)}{\Gamma(2h) \Gamma(1-2h)} e^{i m (\bar \phi_0 + \frac{3 \bar r}{4 \kappa})} f^{lm}_0{}_{\bar \nu},
\label{eqn:FHphiexampleform0}
\ee
with
\bea
f^{lm}_0{}_{\bar \phi} &=& \sum^{\infty}_{N=0} \frac{(i m \frac{3}{2})^{N}}{(N!)^2  \Gamma(1-h-im-N)}    (\log{(i  m \frac{3}{2})}+\psi_0(1-im-h-N)), \\
f^{lm}_0{}_{\bar t} &=& -\frac{\kappa}{im} \sum^{\infty}_{N=0} \frac{(i m \frac{3}{2})^{N} \big(1+(N+h)(\log{(i  m \frac{3}{2})}+\psi_0(1-im-h-N)) \big)}{(N!)^2  \Gamma(1-h-im-N)}    , \\
f^{lm}_0{}_{\bar r} &=& -\frac{1}{2im\bar r}\sum^{\infty}_{N=0} \frac{(i m \frac{3}{2})^{N} \big(1+(N+h+im)(\log{(i  m \frac{3}{2})}+\psi_0(1-im-h-N)) \big)}{(N!)^2  \Gamma(1-h-im-N)}    
\eea
where $\psi_0$ is the digamma function. Just as, for the circular orbit, it turns out that the $l=2$, $m=\pm2$ modes dominate and one has approximately (for convenience choosing $\phi_0$ to eliminate the initial phase of $F^H_{\bar t}$) 
\bea
F^H_{\bar t} &\approx&4.9\times 10^{-3} \frac{q^2 \kappa}{M} \cos{(\frac{3 \bar r}{2 \kappa}+i(1-2h)\log \lambda)} \\
F^H_{\bar r} &\approx& 1.7\times 10^{-3} \frac{q^2}{M \bar r}\cos{(\frac{3 \bar r}{2 \kappa}+i(1-2h)\log \lambda+0.21\pi)} \\
F^H_{\bar \phi} &\approx& 3.0\times 10^{-3} \frac{q^2}{M} \cos{(\frac{3 \bar r}{2 \kappa}+i(1-2h)\log \lambda-0.24\pi)}
\eea

Although it is interesting that this oscillatory behavior is present, in absolute magnitude it can be ignored in general as compared to the inhomogeneous piece. Let us consider the limit $\bar r \to \infty$ but with $\frac{\bar r \lambda}{\kappa} \ll 1$ in order to remain in the near-NHEK approximation. Although it is not obvious to prove, we expect to find the result for circular orbits in NHEK
\be
F^{H}_{\bar \nu}(\bar r \to \infty) \sim \frac{q^2}{\sqrt{3} M} \sum_{lm} \lambda^{2h-1} k_2 [S_{l m}(\frac{\pi}{2})]^2 \cW^{\text{in}}_0   \frac{ \Gamma(1-h-im) \Gamma(h-im)}{\Gamma(2h) \Gamma(1-2h)} (\frac{3im}{4}\frac{\bar r}{\kappa})^{2h-1} e^{i m \bar \phi_0 } f^{lm}_{\infty}{}_{\bar \nu},
\label{eqn:FHphiexampleforminfty}
\ee
with
\bea
f^{lm}_{\infty}{}_{\bar \phi} &=& \cW^{\text{in}}_0, \qquad
f^{lm}_{\infty}{}_{\bar t} = \frac{3 \bar r}{4} \cW^{\text{in}}_0, \qquad
f^{lm}_{\infty}{}_{\bar r} = -\frac{3}{2 \bar r}\cW^{\text{in}}_0{}'.
\eea

The upshot is that the absolute magnitude of the correcting homogenous self-force is small with respect to the Dirichlet self-force, both at $\bar r = 0$ and at $\bar r \rightarrow \infty$. We therefore conjecture that it will remain small throughout the inspiral. Note that even in the case of an excitation of a quasinormal mode (QNM) leading to a gravitational wave burst \cite{nasipak2019repeated}, the difference between the actual QNMs \eqref{eqn:QNM} and the poles associated to the Dirichlet boundary condition will also exhibit the typical suppression $\sim k_2$ such that we still expect a small correction. Parametrically in the high spin limit, the exact result is \eqref{eqn:masternearNHEK}. However in practice, the conclusion of \cite{Hadar:2016vmk} about the formula \eqref{eqn:simpleSF} holds since the traveling waves bring small corrections. 

\section{Summary and prospects}
\label{ccl}

We have investigated the scalar self-force near the horizon of a high spin black hole, at leading order in the high spin limit. We have made concrete the proposal of \cite{Hadar:2016vmk} to first compute the result for circular orbits and subsequently use the relation to more general equatorial orbits as classified in \cite{Compere:2017hsi}. Once the self-force for circular orbits has been computed numerically, the self-force for equatorial orbits can be deduced analytically, up to small corrections due to traveling waves. 

More precisely, we found that the near-horizon geometry and the exterior extremal Kerr geometry do not decouple due to traveling waves in line with previous results \cite{Amsel:2009ev}. We found that it leads to a breaking of conformal symmetry to discrete rescalings. We identified the leading residual discrete scale invariance with associated logarithmic periodicity, which allows to precisely derive the persistent oscillations of the self-force and the associated angular momentum asymptotic flux in the high spin limit \cite{vandeMeent:2016hel}. We then showed that these oscillations are small with respect to the self-force that can be obtained from Dirichlet boundary conditions in the near-horizon region. Applying conformal transformations to obtain general ingoing near-horizon orbits, we showed that the self-force transforms covariantly up to corrections due to traveling waves that are suppressed (with a relative factor of the order of $\sim 10^{-3}$). 

Looking toward the future it would be desirable to extend these results to the gravitational case. The main additional hurdle, which is avoided in the scalar case, is the difficult metric reconstruction. This could in principle be simplified in the near-extremal near-horizon limit, as it is in the static limit, since the metric perturbations are themselves in fact separable \cite{chen2017separating}. In addition, for this gravitational case, it would be interesting to translate the self-force results into invariant quantities, which will simplify the comparison to other self-force results \cite{detweiler2008consequence, akcay2015comparison}. Another interesting avenue for future work would be to consider the implications for cosmic censorship associated with so-called ``deeply bound'' orbits, for which the applied formulation is ideal. Finite size corrections would most likely have to be included \cite{colleoni2015overspinning,colleoni2015self} but they have been computed at leading order in the high spin \cite{Chen:2019hac}. 

\paragraph{Acknowledgements}
K.F. is Aspirant FWO-Vlaanderen (ZKD4846-ASP/18). G.C. is a research associate of the F.R.S.-FNRS and he  acknowledges support from the IISN convention 4.4503.15, the COST Action GWverse CA16104, and FNRS conventions R.M005.19 and J.0036.20.  Y.L. is financially supported
by the China Scholarship Council. This work is also supported by the C16/16/005 grant of the KU Leuven, and by the FWO Grant No. G092617N.

\appendix

\section{Near horizon extremal Kerr}
\label{app:NHEK}

We consider the high spin limit of the nonextremal Kerr black hole with
\bea
\lambda = \sqrt{1-\frac{J^2}{M^4}} \ll 1,
\eea
where $M$, $J$ are respectively the mass and the angular momentum of the rotating black hole. Recall that the Kerr metric in Boyer-Lindquist coordinates is given by 
\bea
ds^2&=&-(1-\frac{2M \hat r}{\Sigma})d\hat t^2+\frac{\Sigma}{\Delta}d\hat r^2+\Sigma d\theta^2 + (\hat r^2+a^2 + \frac{2M a^2 \hat r \sin^2\theta}{\Sigma})\sin^2\theta d\hat \phi^2\nn\\
&&
-\frac{4M a \hat r \sin^2 \theta}{\Sigma} d\hat t d\hat \phi \label{eqn:kerrmetric}
\eea
where $a = J/M$ and
\bea
\Delta \equiv \hat r^2-2M \hat r+a^2,\qquad \Sigma \equiv \hat r^2+a^2 \cos^2\theta. \label{defD}
\eea
We will denote the inner and outer horizons as $\hat r_\pm = M \pm \sqrt{M^2-a^2}$ or $\hat r_\pm = M(1 \pm \lambda)$. As is well known, given a metric depending on a parameter $\lambda$, the limiting spacetime when $\lambda \rightarrow 0$ might depend on the coordinates in which the limit was taken \cite{Geroch:1969ca}. At leading order in the high spin regime, the Kerr metric can be patched with an exterior region isomorphic to the extremal Kerr black hole in Boyer-Lindquist coordinates $(\hat t,\hat r,\theta,\hat \phi)$, the near horizon extremal Kerr geometry in Poincar\'e coordinates $(T,R,\theta,\Phi)$ and the very near horizon extremal Kerr geometry in black hole coordinates $(t,r,\theta,\phi)$. These coordinate systems are related as 
\bea
T &=&\frac{\hat t}{2M}\lambda^{2/3},\qquad R =\frac{\hat r - \hat r_+}{M}\lambda^{-2/3},\qquad \Phi = \hat \phi - \frac{\hat t}{2M},\label{chgt} \\
t &=&\frac{\hat t}{2M}\frac{\lambda}{\kappa},\qquad r =\frac{\kappa }{\lambda} \frac{\hat r - \hat r_+}{M} ,\qquad \phi = \hat \phi - \frac{\hat t}{2M},\label{changecoord}
\eea
where $\kappa$ is arbitrary and therefore factors out of any physical quantity. The NHEK metric \cite{Bardeen:1999px} is
\be
ds^2=2M^2\Gamma(\theta) \left(-R^2 dT^2+\frac{dR^2}{R^2}+d\theta^2+\Lambda^2(\theta)(d\Phi+R dT)^2 \right),
\ee
where the polar functions are 
\be
\Gamma(\theta)=\frac{1+\cos^2\theta}{2},\qquad \Lambda(\theta)=\frac{2\sin\theta}{1+\cos^2\theta}.
\ee
The near-NHEK metric \cite{Amsel:2009ev,Bredberg:2009pv} is
\be
ds^2=2M^2\Gamma(\theta)\left( -r(r+2\kappa) dt^2+\frac{dr^2}{r(r+2\kappa)}+d\theta^2+\Lambda^2(\theta)(d\phi+(r+\kappa) dt)^2 \right).
\ee
It is diffeomorphic to the NHEK metric and in fact, both are part of a global NHEK spacetime. For further details we refer to \cite{Compere:2017hsi,Compere:2018aar}.

\section{Mode-sum Regularization}
\label{app:reg}

Regularization of the self-force entails removing the singular behavior without affecting the motion of the particle. At the level of the Green's function $G_{\text{ret}}(x,x') = G_{R}(x,x') + G_{S}(x,x') $, Detweiler and Whiting asserted that the singular part should have the following properties \cite{Detweiler:2002mi}: $G_S(x,x')$ satisfies the inhomogeneous wave equation, it is symmetric and it vanishes in the chronological past and future of $x$. Taken this for granted, the idea of mode-sum regularization \cite{Barack:1999wf} is to decompose the associated singular field $\Psi^S(x)$ in spherical harmonic  modes $\Psi^S(x) = \sum_{l m}\Psi_{lm}^S Y_{lm}(\theta, \phi)$, each one individually finite, and to obtain the regular field from a mode by mode subtraction of the singular field from the retarded field $\Psi^R = \sum_{l m} (\Psi_{lm}^{\text{ret}}-\Psi_{lm}^S)Y_{lm}(\theta, \phi)$. This procedure can also be performed directly in terms of the force $F_{\mu}$. The relevant retarded field has been computed in the main text while the mode decomposition of the singular field can be found by a local analysis of the Green's function which has been performed previously for a Kerr black hole \cite{Barack:2002mh} and therefore is in principle also already available to us. A caveat is that the values in \cite{Barack:2002mh} are not entirely covariant. We will however cross-check that the infinities are canceled out which justifies \emph{a posteriori} the regularization method, at least for the circular orbits. Let us now turn immediately to our specific setup.

\subsection{Regularization for a generic equatorial Kerr orbit}

In the following, we will keep the polar angle of the trajectory $\theta_0 = \frac{\pi}{2}$ explicit for clarity. Following the mode-sum decomposition, we have 
\bea
F_{\mu} &=& (\sum_{l m} (F^{\text{ret}}_{\mu l m}(R,T)-F^{\text{S}}_{\mu l m}(R,T)) Y_{l m}(\theta,\phi))_{|_z}  \nn \\ 
&=& (\sum_{l} F^{(l) \pm }_{\mu}-A^{\pm}_{\mu}(l+1/2)-B_{\mu}-\frac{C_{\mu}}{l+1/2})-D_{\mu} ,
\eea
with
\bea
F^{(l) \pm }_{\mu} &=& \sum_m \lim_{R\to R(\tau)  \pm \eps} F^{\text{ret}}_{\mu l m}(R,T(\tau)) Y_{l m}(\theta_0,\Phi(\tau)), \nn \\ 
 \lim_{R\to  R(\tau) \pm \eps} \sum_m F^{\text{S}}_{\mu l m}(R,T(\tau))Y_{l m}(\theta_0,\Phi(\tau)) &=& A^{\pm}_{\mu}(l+1/2)+B_{\mu}+\frac{C_{\mu}}{l+1/2}+\sum^{\infty}_{k=2}D^{(k)}_{\mu}(l+1/2)^{-k} \nn ,\\
 \sum^{\infty}_{k=2}D^{(k)}_{\mu}(l+1/2)^{-k} &=& D_{\mu}.
\eea

The regularization parameters $A^{\pm}_{\mu}$, $B_{\mu}$, $C_{\mu}$, and $D_{\mu}$ for a particle on a generic orbit in a Kerr background are given in \cite{Barack:2002mh}\footnote{For us of course $q_s = q$ and $s=0$ but we keep those explicit in order to easily generalize to arbitrary $s$.}
\bea
A^{\pm}_{\hat r} &=& \mp(-1)^{s} \frac{q_s^2}{V}(\frac{\sin^2{\theta_0} g_{\hat r \hat r}}{g_{\theta \theta} g_{\hat \phi \hat \phi}})^{1/2}(V+u^2_{\hat r}/g_{\hat r \hat r})^{1/2};\\
A^{\pm}_{\hat t} &=& -(u^{\hat r}/u^{\hat t})A^{\pm}_{\hat r}; \\
A^{\pm}_{\theta} &=& 0 ;\\
A^{\pm}_{\hat \phi} &=& 0.
\eea
with
\be
V = 1+u^2_{\theta}/g_{\theta \theta}+u^2_{\hat \phi}/g_{\hat \phi \hat \phi}.
\ee
In addition
\be
B_{\mu} = (-1)^{s}\frac{q_s^2}{2 \pi}P_{\mu a b c d}I^{a b c d}
\ee
with $a$, $b$, $\ldots$ going only over the angular coordinates $\theta$, $\hat \phi$ and
\bea
P_{\mu a b c d} &=& (3 P_{\mu a}P_{be}-P_{\mu e} P_{ab})C^{e}_{cd} + \frac{1}{2}(3P_{\mu d}P_{a b c}-(2P_{\mu a b} + P_{ab \mu})P_{cd});\\
P_{\alpha \beta} &=& g_{\alpha \beta}+u_{\alpha}u_{\beta};\\
P_{\alpha \beta \gamma} &=& (u_{\lambda}u_{\gamma}\Gamma^{\lambda}_{\alpha \beta}+g_{\alpha \beta,\gamma}/2);\\
I^{a b c d} &=& (\sin{\theta_0})^{-N}\int^{2 \pi}_{0}G(\gamma)^{-5/2}(\sin{\gamma})^{N}(\cos{\gamma})^{4-N} \d \gamma
\eea
where $N$ is the number of $\hat \phi$ indices. Furthermore
\be
G(\gamma) = \frac{P_{\hat \phi \hat \phi}}{\sin^2{\theta_0}} \sin^2{\gamma} + \frac{2 P_{\theta \hat \phi}}{\sin{\theta_0}} \sin{\gamma} \cos{\gamma}+P_{\theta \theta} \cos^2{\gamma}
\ee
and the only nonvanishing components of $C^{e}_{cd}$ are
\bea
C^{\theta}_{\hat \phi \hat \phi} = \frac{1}{2} \sin{\theta_0} \cos{\theta_0} ,\qquad C^{\hat \phi}_{\theta \hat \phi} = C^{\hat \phi}_{\hat \phi \theta} = -\frac{1}{2} \cot{\theta_0}.
\eea
Finally,
\be
C_{\mu} = D_{\mu} = 0.
\ee

\subsection{Regularization in NHEK}

Taking the NHEK limit, we find
\bea
A^{\pm}_{R} &=& \mp (-1)^{s} \frac{q^2_s}{V} \frac{1}{2 M R }((1+\cos^2{\theta_0})V+R^2u^2_R/M^2)^{1/2},\\
A^{\pm}_{T} &=& -(u^R/u^T)A^{\pm}_{R} ,\\
A^{\pm}_{\theta} &=& A^{\pm}_{\Phi} = 0,
\eea
with
\be
V = 1+ \frac{u_{\theta}^2}{M^2(1+\cos^2{\theta_0})}+\frac{ u^2_{\Phi}}{4 M^2 \sin^2{\theta_0}}(1+\cos^2{\theta_0}).
\ee
In particular, for a circular orbit
\bea
A^{\pm}_{R} &=& \mp  (-1)^s \frac{\sqrt{3}}{4} \frac{q^2_s}{M R_0} ,\\
A^{\pm}_{T} &=& A^{\pm}_{\theta} = A^{\pm}_{\Phi} = 0.
\eea
In addition,
\be
B_{\mu} = (-1)^s\frac{q^2_s}{2 \pi}P_{\mu a b c d}I^{a b c d},
\ee
with\footnote{The indices are now referring to NHEK coordinates. This is important as these expressions are not covariant.}
\bea
P_{\alpha \beta} &=& g_{\alpha \beta}+u_{\alpha}u_{\beta},\\
P_{\alpha \beta \gamma} &=& (u_{\lambda}u_{\gamma}\Gamma^{\lambda}_{\alpha \beta}+g_{\alpha \beta,\gamma}/2),
\eea
and still
\bea
C^{\theta}_{\Phi \Phi} &=& \frac{1}{2} \sin{\theta_0} \cos{\theta_0}, \\
C^{\Phi}_{\theta \Phi} &=& C^{\Phi}_{\Phi \theta} = -\frac{1}{2} \cot{\theta_0},\\
I^{a b c d} &=& (\sin{\theta_0})^{-N}\int^{2 \pi}_{0}G(\gamma)^{-5/2}(\sin{\gamma})^{N}(\cos{\gamma})^{4-N} \d \gamma .
\eea
with $N$ the number of $\Phi$ indices,  
\be
G(\gamma) = \frac{P_{\Phi \Phi}}{\sin^2{\theta_0}} \sin^2{\gamma} + \frac{2 P_{\theta \Phi}}{\sin{\theta_0}} \sin{\gamma} \cos{\gamma}+P_{\theta \theta} \cos^2{\gamma}
	\ee
and 
\bea
P_{\mu a b c d} = (3 P_{\mu a}P_{be}-P_{\mu e} P_{ab})C^{e}_{cd} + \frac{1}{2}(3P_{\mu d}P_{a b c}-(2P_{\mu a b} + P_{ab \mu})P_{cd})
\eea
but now $a$, $b$, $\ldots$ go over $\Phi$ and $\theta$. Note that in particular for the circular orbit on the equatorial plane
\bea
C^{\theta}_{\Phi \Phi} = C^{\Phi}_{\theta \Phi} = C^{\Phi}_{\Phi \theta} = 0.
\eea
Moreover, the only nonzero components of $P_{\mu \nu \gamma}$ are
\bea
P_{T \Phi R} &=& P_{\Phi T R} = P_{R T \Phi} = P_{T R \Phi} = 2 M^2; \\
P_{R R R} &=& - \frac{M^2}{R^3_0} ;\\
P_{T T R} &=& 3 M^2 R_0 ;\\
P_{R \Phi \Phi} &=& P_{\Phi R \Phi} = \frac{8 M^2}{3 R_0}.
\eea
We shall need the following quantities on the worldline 
\bea
P_{\theta \theta} &=& g_{\theta \theta} = M^2, \\
P_{\Phi \Phi} &=& g_{\Phi \Phi} + \frac{4 M^2}{3} = \frac{16}{3} M^2 ,
\eea
such that
\bea
B_{\mu} &=& -(-1)^s M^{-5} \frac{q^2_s}{4 \pi}(2P_{\mu \Phi \Phi} + P_{\Phi \Phi \mu})(P_{\Phi \Phi} I_4+P_{\theta \theta} I_2) \\
&=& -M^{-3}(-1)^s\frac{q^2_s}{4 \pi}(2P_{\mu \Phi \Phi} + P_{\Phi \Phi \mu})(\frac{16}{3} I_4+I_2) .
\eea
Here we defined 
\bea
 I_2  &=& \int^{2 \pi}_0 (\frac{16}{3}\sin^2{\gamma}+\cos^2{\gamma})^{-5/2}(\sin{\gamma})^2 (\cos{\gamma})^2 \d \gamma \\
&=& \frac{3}{676}(19 E(-13/3)-32K(-13/3)),
\eea
and
\bea
 I_4  &=& \int^{2 \pi}_0 (\frac{16}{3}\sin^2{\gamma}+\cos^2{\gamma})^{-5/2}(\sin{\gamma})^4 \d \gamma \\ &=& -\frac{9}{5408}(29 E(-13/3)-120 K(-13/3)),
\eea
where $K$ and $E$ are, respectively, complete elliptic integrals of the first and second kind\footnote{We use the convention of \emph{Mathematica}.}
\bea
K(m) &=& \int^{\pi/2}_{0} (1-m \sin^2{\gamma})^{-1/2} \d \gamma, \\
E(m) &=& \int^{\pi/2}_{0} (1-m \sin^2{\gamma})^{1/2} \d \gamma .
\eea
We conclude 
\bea
B_{R} &=& -M^{-1}(-1)^s\frac{q^2_s}{4 \pi}\frac{16}{3R_0} (\frac{16}{3} I_4+I_2)  \nn\\
      &=& (-1)^s \frac{q_s^2}{\pi M R_0}\frac{3E(-13/3)-16K(-13/3)}{13},\label{NHEKB}\\
B_{\Phi} &=& B_{T} = B_{\theta} = 0,\qquad C_{\mu} = D_{\mu} = 0.
\eea

\subsection{Regularization in near-NHEK}

Taking instead the near-NHEK we find

\bea
A^{\pm}_{r} &=& \mp(-1)^{s} \frac{q^2_s}{V}(\frac{\sin^2{\theta_0} g_{r r}}{g_{\theta \theta} g_{\phi \phi}})^{1/2}(V+u^2_{r}/g_{r r})^{1/2},\\
A^{\pm}_{t} &=& -(u^{r}/u^{t})A^{\pm}_{r}, \qquad A^{\pm}_{\theta} = 0, \qquad A^{\pm}_{\hat \phi} = 0,
\eea
with
\be
V = 1+u^2_{\theta}/g_{\theta \theta}+u^2_{ \phi}/g_{\phi \phi}.
\ee

For a circular orbit we find more explicitly

\bea
A^{\pm}_{r} &=& \mp(-1)^{s} \frac{q^2_s}{M r_0}, \frac{\sqrt{3+6\kappa_0-\kappa^2_0}}{4(1+2\kappa_0)},\\
A^{\pm}_{t} &=& 0,\qquad A^{\pm}_{\theta} = 0, \qquad A^{\pm}_{\hat \phi} = 0.
\eea

To compute $B_{\mu}$, note

\bea
P_{t \phi r} &=& P_{\phi t r} = 2M^2, \\
P_{t r \phi} &=& P_{r t \phi} = 6M^2 \frac{(1+\kappa_0)^2}{3+6\kappa_0-\kappa_0^2}, \\
P_{r r r} &=& -\frac{M^2}{r_0^3} \frac{(1+\kappa_0)}{(1+2\kappa_0)^2},\\
P_{t t r} &=& 3 M^2 r_0 (1+\kappa_0), \\
P_{r \phi \phi} &=& P_{\phi r \phi}= \frac{8 M^2}{r_0} \frac{1+\kappa_0}{3+6\kappa_0-\kappa^2_0}, \\
P_{t r t} &=& P_{r t t} = 6 M^2 r_0 \kappa_0^2 \frac{1+\kappa_0}{3+6\kappa_0-\kappa_0^2}, \\
P_{\phi r t} &=& P_{r \phi t} = 8 M^2  \kappa_0^2 \frac{1}{3+6\kappa_0-\kappa_0^2}, 
\eea

and

\bea
P_{\theta \theta} &=& M^2, \\
P_{\phi \phi} &=& \frac{16 M^2(1+2\kappa_0)}{3+6\kappa_0-\kappa^2_0}, \\
P_{\theta \phi} &=& 0.
\eea
Therefore, we can write

\be
B_{\mu} = -(-1)^s \frac{q^2_s }{4 \pi M^3} (2 P_{\mu \phi \phi} + P_{\phi \phi \mu})(\tilde{\kappa} \tilde{I}_4 + \tilde{I}_2)
\ee
where
\bea
 \tilde{I}_2 &=& \int^{2 \pi}_0 (\tilde{\kappa}\sin^2{\gamma}+\cos^2{\gamma})^{-5/2}\sin^2{\gamma}\cos^2{\gamma} \d \gamma  \\
&=& \frac{4}{3(1-\tilde{\kappa})^2}\left( (1+\frac{1}{\tilde{\kappa}})E(1-\tilde{\kappa}) -2 K(1-\tilde{\kappa}) \right), \nn\\
 \tilde{I}_4 &=& \int^{2 \pi}_0 (\tilde{\kappa}\sin^2{\gamma}+\cos^2{\gamma})^{-5/2}\sin^4{\gamma} \d \gamma  \\
&=& -\frac{4}{3(1-\tilde{\kappa})^2\tilde{\kappa}^2}\Big( 2(-1+2\tilde{\kappa})E(1-\tilde{\kappa})+(1-3\tilde{\kappa})K(1-\tilde{\kappa}) \tilde{\kappa}\Big),
\eea
where we defined 
\bea
\tilde{\kappa}&=&\frac{16(1+2\kappa_0)}{3+6\kappa_0-\kappa^2_0}.
\eea
We have finally
\bea
B_r &=&  -(-1)^s \frac{q^2_s}{4 \pi M r_0} \frac{16(1+\kappa_0)}{3+6\kappa_0-\kappa^2_0} (\tilde{\kappa}\tilde{I}_4 + \tilde{I}_2)\\
&=&  -(-1)^s \frac{q^2_s}{\pi M r_0} \frac{16(1+\kappa_0)}{3+6\kappa_0-\kappa^2_0} \frac{E(1-\tilde \kappa) - \tilde \kappa K(1-\tilde \kappa)}{\tilde \kappa (1-\tilde \kappa)},\\
B_{\phi} &=& B_{\theta} = B_t = 0 ,\qquad C_{\mu} = D_{\mu} = 0.
\eea
For $\kappa_0 = 0$, we recover the NHEK result \eqref{NHEKB} with $R_0=r_0$. 

\section{$F^H$ for equatorial orbits}
\label{app:Bs}

In Sec. \ref{sec:genhom}, we have written the additionally required ingoing modes, needed to satisfy the appropriate boundary conditions in asymptotically flat space. Here, we make explicit which $B_{\hat lm\Omega}$ or $B_{\hat lm\omega}$ is required for each of the families of equatorial orbits described in \cite{Compere:2017hsi}. First, we define
\bea
B_{\text{circ}} &\equiv&   \frac{q S_{\hat lm}(\pi/2)}{\sqrt{3} M}  \frac{\Gamma(h-im)}{im \Gamma(2 h)}  \cW^{\text{in}}(R_0), \\
B^{\text{near}}_{\text{circ}} &\equiv& - \frac{q r_0 \sqrt{3+6\kappa_0-\kappa^2_0}}{4 M \kappa_0}   \frac{S_{\hat l m}(\frac{\pi}{2}) \Gamma(h-i m) \Gamma(h- \frac{i m \tilde \omega}{\kappa})}{(1-2h)\Gamma(2h-1) \Gamma(1-i m(1 + \frac{\tilde\omega}{\kappa}))} \cR^{\text{in}}(r_0). 
\eea
such that the appropriate values of $B_{\hat lm\Omega}$ for the NHEK circular orbit is given by
\be
B_{\hat l m \Omega}(\gamma \rightarrow \text{Circular}_*) =\sqrt{2 \pi} \delta(\Omega-m\tilde{\Omega}) B_{\text{circ}},
\ee
while for the near-NHEK circular orbit it is
\be
B_{\hat l m \Omega}(\gamma \rightarrow \text{Circular}(\ell ) ) =\sqrt{2 \pi} \delta(\omega-m\tilde{\omega}) B^{\text{near}}_{\text{circ}}.
\ee
For more general orbits, using inverse Laplace transforms or inferring the $s=0$ case from \cite{Compere:2017hsi}, we obtain
	\begin{itemize}
	
	\item $\gamma \rightarrow $ Plunging$_*(E)$
	
	\be
	B_{\hat lm\Omega} = 2\frac{B_{\text{circ}}}{\sqrt{2 \pi}} \frac{\sqrt{-2im\tilde{\Omega}}}{\sqrt{-2i\Omega}}K_{1-2h}(2\sqrt{-m \Omega \tilde{\Omega}}),
	\ee
	
	with $R_0 = \frac{2 \ell_*}{E}$. This expression matches an unevaluated integral expression given in Section 2 of \cite{Hadar:2015xpa}.
	
	\item $\gamma \rightarrow $ Plunging$(E,\ell)$
	
	\be
	B_{\hat lm\Omega} = \frac{B^{\text{near}}_{\text{circ}}}{\sqrt{2 \pi}} (2\kappa)^{-h}(-i\Omega)^{-1}\Gamma(1-h-\frac{i m \tilde{\omega}}{\kappa})(\cot{\frac{\zeta}{2}})^{-\frac{i m \tilde{\omega}}{\kappa}} e^{-i\Omega \cot{\zeta}}W_{\frac{\i m \tilde{\omega}}{2},\frac{1}{2}-h}(\frac{-2 i \Omega}{\sin{\zeta}}),
	\ee
	
	with $\kappa_0 = (\frac{2 \ell}{\sqrt{3(\ell^2-\ell^2_*)}}-1)^{-1}$, $E = \frac{\sqrt{3(\ell^2-\ell^2_*)}}{2}(\sin{\zeta}+T_0(\cos{\zeta}-1))$, $\bar{T_0} = \frac{-\cos{\zeta}+T_0 \sin{\zeta}}{\sin{\zeta}+T_0(\cos{\zeta}-1)}$ \footnote{$T_0$ is simply an intermediate parameter, $\bar{T}_0$ is the meaningful orbital parameter in the physical spacetime.}.
	
	\item $\gamma \rightarrow $ Marginal$(\ell)$
	
	\be
	B_{\hat lm\Omega} = \frac{B^{\text{near}}_{\text{circ}}}{\sqrt{2 \pi}} (2\kappa)^{-h}(-i\Omega)^{-1+\frac{im\tilde{\omega}}{\kappa}}\Gamma(1-h-\frac{i m \tilde{\omega}}{\kappa}),
	\ee
	
	with $\kappa_0 = (\frac{2 \ell}{\sqrt{3(\ell^2-\ell^2_*)}}-1)^{-1}$.
	
	\item $\gamma \rightarrow $  Plunging$_*(e=0)$
	
	\be
	B_{\hat lm\omega} = 2 (2 \kappa)^{-1}\frac{B_{\text{circ}}}{\sqrt{2 \pi}} (-im\tilde{\Omega})^{i \frac{\omega}{\kappa}} \Gamma(h-i\frac{\omega}{\kappa}),
	\label{eqn:Bplungee0}
	\ee
	
	with $R_0 = \frac{e^{\kappa \bar{t}_0}}{\kappa}$. This exactly matches (4.18) of \cite{Hadar:2014dpa}.
	
	\item $\gamma \rightarrow $  Plunging$_*(e)$
	
	\be
	B_{\hat lm\omega} = 2 (2 \kappa)^{-1}\frac{B_{\text{circ}}}{\sqrt{2 \pi}} \Gamma(h-i\frac{\omega}{\kappa})(-\tan{\frac{\zeta}{2}})^{\frac{i \omega}{\kappa}} e^{im \tilde{\Omega} \cot{\zeta}} W_{\frac{i \omega}{\kappa},h-\frac{1}{2}}(2 im \tilde{\Omega} \csc{\zeta}),
	\ee
	
	with $R_0 = \frac{e^{\kappa t_0}}{\kappa}$, $e = \kappa^2 \ell_* \sin{\zeta} e^{-\kappa t_0}$ and $\bar{t}_0 = \frac{i \pi}{\kappa} + \frac{1}{\kappa} \log{\frac{\sin{\zeta}}{1+\cos{\zeta}}}$. This matches an unevaluated integral expression given in Sec. 3 of \cite{Hadar:2015xpa}. 
	
	\item $\gamma \rightarrow $ Plunging$(e, \ell)$
	
	\bea
	B_{\hat lm\omega} &=& \frac{B^{\text{near}}_{\text{circ}}}{\sqrt{2 \pi}} \kappa^{-1-h} (1+\chi)^{\frac{i m \tilde{\omega}}{\kappa}-h} B(h-\frac{i\omega}{\kappa},1-h-\frac{i m \tilde{\omega}}{\kappa}) \nn \\ &\times&	{}_2F_1(h - \frac{i m \tilde{\omega}}{\kappa}, h - \frac{i \omega}{\kappa},1 - \frac{i m \tilde{\omega}}{\kappa}- \frac{i \omega}{\kappa},- \frac{1-\chi}{1+\chi}),
	\eea
	
	with $e = \frac{1}{2}\sqrt{3(\ell^2-\ell^2_*)} \kappa \chi$, $\bar{t}_0 = -\frac{1}{2 \kappa} \log{\frac{1+\chi}{1-\chi}}$. This matches an implicit expression given in \cite{Hadar:2016vmk}. 
	
\end{itemize}

\newpage

\section{Detailed numerical results}
\label{app:num}

\begin{figure}[!hbt]
	\centering
	\subfigure{
		\includegraphics[width=.48\textwidth]{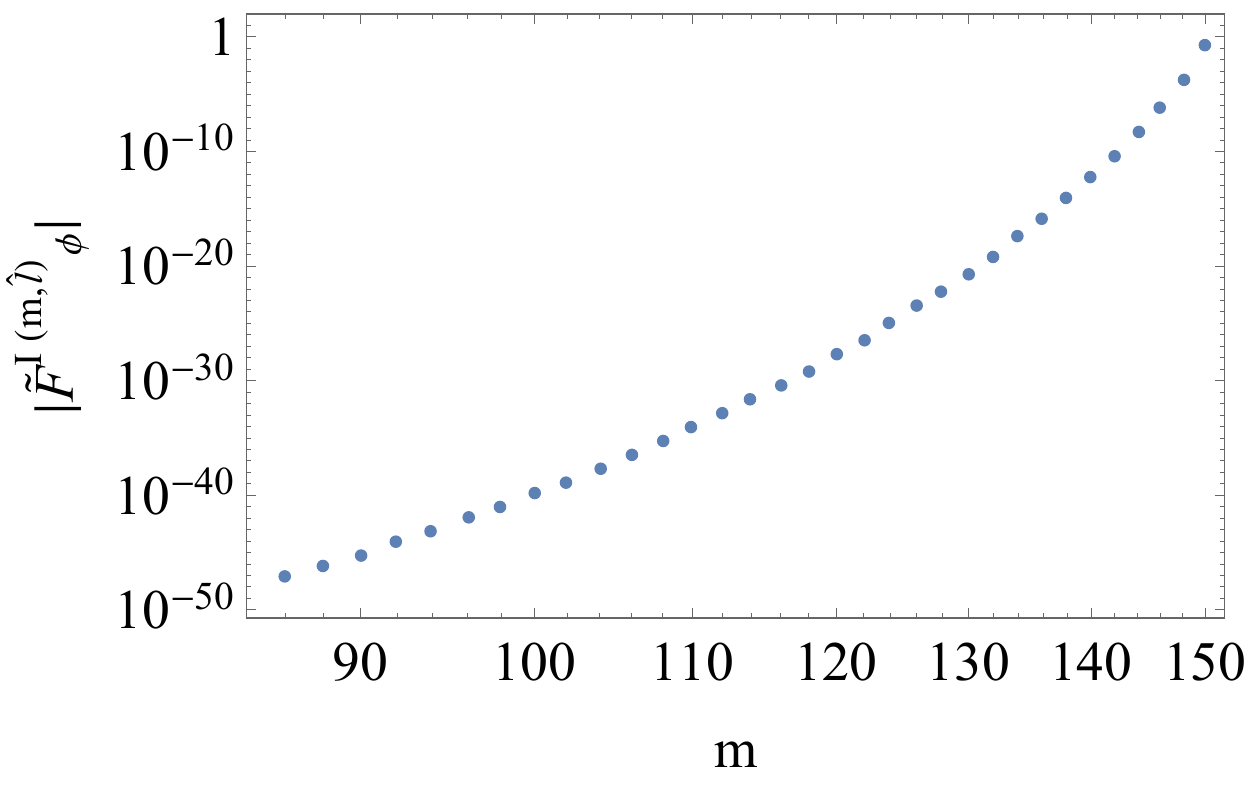}}
	\caption{Individual $m$ mode contributions to $\tilde F_{\phi}^I{}^{(\hat l = 150)} (\kappa_0=6.289)$.}
	\label{fig:Fphimcontr}\vspace{-10pt}
\end{figure}

In this appendix we provide the numerically computed values for the self-force with Dirichlet boundary conditions as displayed on Fig. \ref{fig:FinhomnearNHEK}. We also provide an ancillary file with this data. The method used to derive these values was explained in Sec. \ref{sec:nearNHEKSF}. We can compute only up to a finite $\hat l$ (respectively $l$) but we can give on estimate of the ``tail'' contribution, beyond the cutoff $\hat l_{\text{cutoff}}$, by 
\be
\epsilon_{\text{trunc}} = \sum_{\hat l_{\text{cutoff}}+1}^{\infty} a_0 e^{-a_1 \hat l}
\ee
for $\tilde{F}_{\phi}^{(\text{tail})}$ and
\be
\epsilon_{\text{trunc}} = \sum_{l_{\text{cutoff}}+1}^{\infty} \frac{a_2}{l^2}
\ee
for $\tilde{F}_{r}^{(\text{tail})}$. The coefficients $a_0$, $a_1$, $a_2$ are found from a fit of the highest explicitly computed data points. Especially for $\tilde{F}_{\phi}^{(\text{tail})}$ at low $\kappa_0$ this truncation error can be made very small. Nearing the lightring $\kappa_{l.r.}$ on the other hand the expected tail behavior sets at higher $\hat l$ ($l$) and it becomes prohibitively difficult to get accurate results or, in particular, a faithful estimate of the tail contribution. The results given for such $\kappa_0$ values (say $\kappa_0 > 5$) are therefore better considered to give an order of magnitude estimate. In fact to compute $\tilde{F}^I_{\phi}$ for $\kappa_0 > 6.075$ (i.e., last three entries in Table \ref{tbl:nearFtildephi}) we have used extrapolations of $h_{lm}$ and $S_{lm}(\theta)$ to high $l$. In addition, for these computations, we have neglected the contribution for modes of low $m$. We illustrate on Fig. \ref{fig:Fphimcontr} that this is a reasonable approximation. Due to this approximation, the extrapolation of $h_{lm}$ corresponds in particular to the double scaling limit discussed in \cite{Hod:2015cqa} which we found to be consistent. For the last entry $\kappa_0 = 6.464$, we have also used interpolation methods at large $\hat l$ rather than computing each $\hat l$ contribution explicitly. We therefore stress that these entries in particular should only be taken as order of magnitude estimates. To similarly extend the results for $\tilde{F}^I_{r}$ to higher $\kappa_0$ with confidence proved beyond our computational capabilities at present.

\begin{table}
	\caption{Numerical results for $\tilde{F}^I_{\phi}(\kappa_0)$. The error for $\tilde{F}^I_{\phi}$ is $\pm 0.5\times 10^{-3}$ until $\epsilon_{\text{trunc}}$ becomes of this order or larger in which case it is itself an estimate for the error. The asterisk (*) indicates the use of extrapolation methods and stresses that these results should be considered more tentative.} 
	\label{tbl:nearFtildephi}
	\begin{center}
		\begin{tabular}{ |l c c r|l c c r| }
			\hline
			$\kappa_0$ & $\tilde{F}^I_{\phi}$ &  $\epsilon_{\text{trunc}}$ & $\hat l_{\text{cutoff}}$ & $\kappa_0$ & $\tilde{F}^I_{\phi}$ &  $\epsilon_{\text{trunc}}$ &  $\hat l_{\text{cutoff}}$ \\
			\hline
			$0.000$ & $-0.225$ &  $-10^{-10}$ & 100 & $3.499$ & $-1.270$ &  $-10^{-7}$ & 150 \\
			$0.044$ & $-0.225$ &  $-10^{-10}$ & 100 & $4.025$ & $-1.847$ &  $-10^{-5}$ & 150 \\
			$0.175$ & $-0.231$ &  $-10^{-10}$ & 100 & $4.530$ & $-2.814$ &  $-0.0001$ & 150 \\
			$0.390$ & $-0.247$ &  $-10^{-10}$ & 100 & $5.000$ & $-4.553$ &  $-0.002$ & 150\\
			$0.682$ & $-0.278$ &  $-10^{-9}$ & 100 & $5.421$ & $-7.97$ &  $-0.03$ & 150 \\
			$1.043$ & $-0.327$ &  $-10^{-9}$ & 100 & $5.783$ & $-15.8$ &  $-0.1$ & 200 \\
			$1.464$ & $-0.402$ &  $-10^{-8}$ & 100& $6.075$ & $-36$ &  $-2$ & 200 \\
			$1.934$ & $-0.511$ &  $-10^{-7}$ & 100 & $6.289$ & $-125$ &  $-4$ & 500* \\
			$2.439$ & $-0.670$ &  $-10^{-6}$ & 100 & $6.420$ & $-8\times 10^2$ &  $-3 \times 10^2$ & 1000* \\
			$2.965$ & $-0.907$ &  $-10^{-5}$ & 100 & $6.464$ & $-4\times10^6$ &  $-9\times 10^6$ & 250 000** \\
			\hline
		\end{tabular}
	\end{center}\vspace{-15pt}
\end{table}

\begin{table}
	\caption{Numerical results for $\tilde{F}^I_{r}(\kappa_0)$. The error for $\tilde{F}^I_{r}$ is $\pm 0.5\times 10^{-3}$ until $\epsilon_{\text{trunc}}$ becomes of this order or larger in which case it is itself an estimate for the error. Remark however, that it also becomes more uncertain itself when $\kappa_0 \to \kappa_{l.r.}$ as the clean $l^{-2}$ tail moves beyond our reach. } 
	\label{tbl:nearFtilder}
	\begin{center}
	\begin{tabular}{ |l c c r|l c c r| }
		\hline
		$\kappa_0$ & $\tilde{F}^I_{r}$ &  $\epsilon_{\text{trunc}}$ & $l_{\text{cutoff}}$ & $\kappa_0$ & $\tilde{F}^I_{r}$ &  $\epsilon_{\text{trunc}}$ & $l_{\text{cutoff}}$ \\
		\hline
		$0.000$ & $-0.212$ &  $0.009$ & $80$ &  $3.499$ & $-0.114$ &  $0.008$ & $130$ \\
		$0.044$ & $-0.204$ &  $0.009$ & $80$ & $4.025$ & $-0.122$ &  $0.010$ & $140$ \\
		$0.175$ & $-0.183$ &  $0.007$ & $80$ & $4.530$ & $-0.136$ &  $0.013$ & $150$ \\
		$0.390$ & $-0.156$ &  $0.004$ & $130$ & $5.000$ & $-0.16$ &  $0.02$ & $160$ \\
		$0.682$ & $-0.136$ &  $0.004$ & $130$ & $5.421$ & $-0.19$ &  $0.04$ & $160$\\
		$1.043$ & $-0.124$ &  $0.007$ & $80$ & $5.783$ & $-0.26$ &  $0.08$ & $170$ \\
		$1.464$ & $-0.114$ &  $0.006$ & $90$ & &  &   &\\
		$1.934$ & $-0.109$ &  $0.006$ &  $100$ & &  & & \\
		$2.439$ & $-0.108$ &  $0.006$ &  $110$ & & &  & \\
		$2.965$ & $-0.109$ &  $0.007$ &  $120$ & &  &   & \\
		\hline
	\end{tabular}
\end{center}\vspace{-15pt}
\end{table}

\newpage


\providecommand{\href}[2]{#2}\begingroup\raggedright\endgroup

\end{document}